\documentclass[%
 reprint,
 nofootinbib,
 amsmath,amssymb,
 aps,
]{revtex4-2}

\usepackage{graphicx}
\usepackage{dcolumn}
\usepackage{bm}

\usepackage{color}

\begin{document}

\preprint{APS/123-QED}

\title{
        General Relativistic Effects on Hill Stability of Multi-Body Systems I: \\ 
        Stability of Three-Body Systems Containing a Massive Black Hole
      }

\author{Haruka Suzuki}
\email{suzuki@heap.phys.waseda.ac.jp}
\affiliation{%
Graduate School of Advanced Science and Engineering, \\ Waseda University, Shinjuku, Tokyo 169-8555, Japan
}%
\author{Yusuke Nakamura}%
\email{yusuke.nakamura@mae.nagoya-u.ac.jp}
\affiliation{%
Department of Aerospace Engineering, \\ Nagoya University, Furo-cho, Chikusa Nagoya 474-8603, Japan
}%
\author{Shoichi Yamada}%
\email{shoichi@waseda.jp}
\affiliation{%
Advanced research Institute for Svience and Engineering, \\ Waseda University, 3-4-1 Okubo, Shinjuku, Tokyo 169-8555, Japan
}%

\date{\today}

\begin{abstract}
We study the effects of general relativistic gravity on the Hill stability, that is, the stability of a multi-body system against a close approach of one orbit to another, which 
has been hitherto studied mainly in Newtonian mechanics and applied to planetary systems.
We focus in this paper on the three-body problem and extend the Newtonian analyses to the general relativistic regime in the post-Newtonian approximation. 
The approximate sufficient condition for the relativistic Hill stability of three-body systems is derived analytically and its validity and usefulness are confirmed numerically.
In fact, relativity makes the system more unstable than Newtonian mechanics in the sense of the Hill stability as expected by our theoretical prediction.
The criterion will be useful 
to analyze the results of large-scale N-body simulations of dense environments, in which the stability of three-body sub-systems is important.
\end{abstract}
\maketitle

\section{\label{sec:Intro}Introduction}
The orbital stability of multi-body systems is one of the oldest research fields in astronomy.
Numerous astronomers, physicists and mathematicians have tackled this problem multifacetedly.
The Hill stability problem is one of the research topics in the field, started from the study on the lunar motion by \citet{Hill1878}. 
The Hill stability is an orbital stability against a close approach: the system is said to be Hill stable if none of the pairs of orbits in the system experiences a close approach for all the time.

Hill's paper \cite{Hill1878} and the following works \cite{Szebehely67, Henon70, Henon86} analyzed the Hill stability for limited three-body systems, using Jacobi integral.
These systems are called the circular restricted three-body systems, in which two components have much smaller masses than the other one and are orbiting this massive component in a coplanar and circular way.
They found for this class of three-body systems that if the initial distance between the two orbits $\Delta$ is large enough, the lighter two objects are separated by the so-called forbidden region for all the time and hence cannot come close to each other, i.e., the system is Hill stable.
A more detailed analysis using Hill's coordinates was given by \citet{Henon86}. 

Various authors have extended these investigations to more general three-body systems (see \cite{Marchal82, Milani83a, Roy84} and references therein).
The essential idea in these works is that the allowed and forbidden regions for each component in the system can be analyzed from the relation between the values of the total energy and angular momentum.

The Hill stability problem has been also investigated in the context of the evolution and formation of planetary systems.
After the first discovery of the extra-solar planetary system in 1992 \cite{Wolszczan92}, \citet{Gladman93} recast the sufficient condition for the Hill stability derived by \citet{Marchal82} into simple inequalities for the orbital separations $\Delta > \Delta_\mathrm{cr}$ by employing several approximations appropriate for the planetary systems.

In 1996, \citet{Chambers96} explored the Hill stability for four- and more-than-four-body systems numerically.
Quite unexpectedly, the sufficient condition for the Hill stability of a similar sort was not found for these more-than-three-body systems. 
Instead, \citet{Chambers96} obtained a log-linear relation between the time it takes the system to experience a close approach $T_\mathrm{stab}$ and the initial orbital separation $\Delta$.
The configurations considered in their paper were again limited to those with three components that have small masses and rotate around a massive object in coplanar and circular orbits.
Numerous authors have followed suit and investigated the relation between $T_\mathrm{stab}$ and $\Delta$ for other systems with different configurations:
elliptical orbits \cite{Ito99, Chatterjee08, Smith09, Pu15}, noncoplanar orbits \cite{Marzari02}, unequal initial orbital separations \cite{Marzari14}, and somewhat more massive planets \cite{Morrison16}.
The resultant relations between the system scale and the onset time of instability have been applied to the studies of formation of planetary systems \cite{Chambers98, Iwasaki06, Zhou07}, 
in which the instability time is supposed to give the timescale for the collision of planetesimals.

So far, almost all works discussing the Hill stability have used Newtonian mechanics.
It is fine for the studies of satellites, planets and planetesimals.
It is not so fine, however, if one wants to consider multi-body systems composed of compact objects such as black holes (BH), neutron stars (NS) and white dwarfs in tight orbits. 
General relativity (GR) must be taken into account then instead of Newtonian mechanics to calculate the evolution of such systems.

Although such relativistic multi-body systems may not be as common as the Newtonian systems in our universe, there is indeed an example actually observed:
the PSR J0337+1715 system is a relativistic three-body system composed of a millisecond pulsar and two white dwarfs \cite{Ransom14}.
More relativistic systems containing massive BHs will be detected with gravitational waves (GW) by future satellite-borne GW detectors like Laser Interferometer Space Antenna (LISA) \cite{Randall_2019, Hoang19, Gupta19} or with radio observations of pulsars \cite{Suzuki19, Suzuki20}. 
In these systems, multi-body interactions between compact objects are definitely important to make compact binaries that experience coalescence in the Hubble time \cite{Samsing_2014, Leigh16, Leigh17, Liu17, Zevin19}. 
The multi-body interactions in dense environments like globular clusters or galactic centers are investigated with large-scale N-body numerical simulations \cite{Secunda19, Fragione_Antonini19}, in which the effect of the presence of supermassive black holes (SMBH) or intermediate-mass black holes (IMBH) at the center of the system is also explored \cite{Trani19a, Fragione_Bromberg19, Trani19b}. 

In spite of the increasing attention to the relativistic multi-body systems, few researches have been devoted to a systematic examination of the stability of such systems in general relativity.
One exception is the paper by \citet{Ge91}, 
which was limited to the Schwarzschild geometry, however, and the application of their analysis to other systems with various configurations is not easy.

Our motivation in this paper is hence to investigate the GR effect on the Hill stability.
We use the post-Newtonian approximation instead of the fully relativistic gravity as in \citet{Ge91} to facilitate the application to different configurations. 
In this paper, we address only the Hill stability problem for relativistic three-body systems and confine the discussion to the configurations that have an SMBH or IMBH at the center of the system and two much-smaller-mass objects orbiting it, for simplicity.
Other configurations and more-than-three-body systems will be discussed in our subsequent papers.
We extend the theoretical Newtonian analysis in the previous works to the post-Newtonian gravity and give approximate sufficient conditions for the relativistic Hill stability.
Numerical simulations are also conducted in the post-Newtonian approximation to test the conditions.
We demonstrate that the systems are more Hill-unstable in the relativistic calculation than in the Newtonian calculation and that the results are quantitatively in agreement with our theoretical prediction.
Our conditions will be useful not only to predict the stability of relativistic three-body systems but also to analyze the results of large-scale N-body simulations of dense star clusters.

This paper is organized as follows.
In  \S\ref{sec:Theory}, we explain the Hill stability more precisely.
The theoretical analysis to give the approximate sufficient conditions for the relativistic Hill stability, one of the main achievements of this paper, is also given in this section.
In \S\ref{sec:Num}, we describe the method of the numerical simulations run in this paper to test the stability conditions. 
The results and some discussions are presented in \S\ref{sec:Result}.
We conclude the paper in \S\ref{sec:Conclusion}.

\section{\label{sec:Theory}Theoretical Analysis of Hill Stability}

\subsection{\label{subsec:Hill_N} Brief review of Newtonian Analysis}

The generalized Hill stability for three-body systems is defined by \citet{Marchal82} as follows:
a triple system is Hill-stable if it can be grouped into a close bounded binary and a third body orbiting it.
Note that the case, in which the third body escapes from the system, is Hill-stable according to this definition.

The Hill stability of a given three-body system can be judged with the topological analyses that were much elaborated in 1970's and 1980's (see e.g. \cite{Henon70, Henon86, Marchal82, Milani83a, Roy84}).
The phase space of the third body in the triple system is divided into allowed and forbidden regions.
If the orbit of the third body is separated from the orbit of the inner binary by the forbidden region in the phase space, they cannot approach each other closely and the system is Hill-stable. 
This means that the existence of the forbidden region between the two orbits is the sufficient condition for the Hill stability of the three-body system.
Below we summarize how the allowed and forbidden regions are obtained from the quantities that characterize the three-body system in Newtonian mechanics.
Although the contents in this subsection are not original, it will facilitate the understanding of our extensions that follow in later sections.

In the analysis of the three-body system, Sundman's inequality (see e.g. \cite{Ge91, Ge94}), which is written as
    \begin{eqnarray}
        &&\left( \sum_j m_j r_j^2 \right) \left( \sum_j m_j v_j^2 \right) \nonumber \\
                && \ge  \left| \sum_j m_j \bm{r}_j \times \bm{v}_j \right|^2 + \left| \sum_j m_j \bm{r}_j \cdot \bm{v}_j \right|^2 ,
        \label{eq:Sundman}
    \end{eqnarray}
is known to be very convenient.
In this inequality, $m_j$, $\bm{r}_j$ and $\bm{v}_j$ normally mean the mass, position and velocity vectors of the $j$-th object, respectively.
We remark, however, that the two vectors 
are arbitrary actually and $m_j$ can take an arbitrary positive value in fact.
The subscript $j$ runs from 1 to $N$, which is an arbitrary integer. 
In this paper we take $N=3$.
The proof of this inequality is given in Appendix \ref{app:Sundman}.
If $m_j$, $\bm{r}_j$ and $\bm{v}_j$ are chosen to be the mass, position and velocity vectors as usual, inequality \eqref{eq:Sundman} can be rewritten with some characteristic quantities of the system as
    \begin{equation}
        2\left( \sum_j m_j r_j^2 \right) \left(  \mathcal{H}_\mathrm{N} - U \right) \ge J^2 + \left| \sum_j m_j \bm{r}_j \cdot \bm{v}_j \right|^2,
        \label{eq:Sundman2}
    \end{equation}
where $\mathcal{H}_\mathrm{N}$, $U$ and $J$ are the total Hamiltonian, the gravitational potential and the magnitude of the total angular momentum, respectively, and are given as
    \begin{equation}
        \mathcal{H}_\mathrm{N}= \frac{1}{2} \sum_j m_j v_j^2 + U; 
    \end{equation}
    \begin{equation}
        U = -\frac{1}{2} \sum_i \sum_{j \neq i} \frac{G m_i m_j}{r_{ij}},
        \label{eq:N_potential}
    \end{equation}
where $G$ is the gravitational constant and $r_{ij}=|\bm{r}_i-\bm{r}_j|$ is the distance between the $i$-th and $j$-th objects, and
    \begin{equation}
        J = \left|\sum_j m_j \bm{r}_j \times \bm{v}_j \right|.
    \end{equation}
Since the second term on the right-hand side of inequality~\eqref{eq:Sundman2} is positive, we have
    \begin{equation}
        2\left( \sum_j m_j r_j^2 \right) \left(  \mathcal{H}_\mathrm{N} - U \right) \ge J^2 .
        \label{eq:Sundman3}
    \end{equation}
This inequality depends only on the positions.
With two of the three positions being fixed, inequality~\eqref{eq:Sundman3} gives the condition that the remaining position should satisfy, which then provides the allowed regions 
 characterized by the conserved quantities of the system $\mathcal{H}_\mathrm{N}$ and $J$.
The Hill stability of the three-body system can be hence judged from 
the conserved quantities and the positions of two objects in the system.

\citet{Marchal82} parameterized inequality~\eqref{eq:Sundman3} in a nice way and gave the sufficient condition of the Hill stability for general three-body systems as an inequality.
\citet{Gladman93} rewrote approximately the inequality in an even simpler form with the orbital elements when the mass of the central object overwhelms other objects orbiting it and the orbital planes are coplanar.
In that limited case, the conserved quantities are approximately given with the initial orbital elements as
    \begin{eqnarray}
        \mathcal{H}_\mathrm{N} &\approx& -\frac{G m_1 m_2}{2a_\mathrm{in}} -\frac{G m_1 m_3}{2a_\mathrm{out}}, 
        \label{eq:H_N}
        \\
        J^2 &\approx& (J_\mathrm{in} + J_\mathrm{out})^2 ,
        \label{eq:J^2_N}
    \end{eqnarray}
where $a$ and $e$ are the semi-major axis and the eccentricity, and the subscripts 'in' and 'out' mean the inner and outer orbits, respectively; 
$J_\mathrm{in}$ and $J_\mathrm{out}$ are defined as
    \begin{equation}
        J_\mathrm{in} = \sqrt{G \frac{m_1^2 m_2^2}{m_1+m_2} a_\mathrm{in}(1-e_\mathrm{in}^2) },
        \label{eq:J_in}
    \end{equation}
    \begin{equation}
        J_\mathrm{out} = \sqrt{G \frac{m_1^2 m_3^2}{m_1+m_3} a_\mathrm{out}(1-e_\mathrm{out}^2) }.
        \label{eq:J_out}
    \end{equation}
When the two orbiting objects have equal masses,
the sufficient conditions obtained in his work are summarized as follows:
    \begin{enumerate}
        \item for initially circular orbits ($e_\mathrm{in}, e_\mathrm{out}=0$)
            \begin{equation}
                \frac{ a_\mathrm{out}-a_\mathrm{in}}{ a_\mathrm{in}} > 3 \mu^\frac{1}{3} ,
                \label{eq:Gladman_c}
            \end{equation}
            with $\mu$ being the ratio of the mass of the orbiting objects to that of the central object,
        \item for initially low eccentric orbits ($e_\mathrm{in}, e_\mathrm{out} \leq \mu^\frac{1}{3}$)
            \begin{equation}
                \frac{ a_\mathrm{out}-a_\mathrm{in}}{ a_\mathrm{in}} 
                    > \sqrt{\frac{8}{3} (e_\mathrm{in}^2 + e_\mathrm{out}^2) + 9\mu^\frac{2}{3} },
                \label{eq:Gladman_el}
            \end{equation}
        \item for initially highly eccentric orbits ($e_\mathrm{in}=e_\mathrm{out}=e > \mu^\frac{1}{3}$)
            \begin{eqnarray}  
                \frac{ a_\mathrm{out}-a_\mathrm{in}}{ a_\mathrm{in}}  &>&
                    \Bigg(
                        \sqrt{ \frac{3+e^2}{2(1-e^2)}-\frac{1}{2} \sqrt{\frac{9-e^2}{1-e^2}} } 
                \nonumber \\
                &&        + \frac{1}{2} \sqrt{\frac{9-e^2}{1-e^2}} -\frac{1}{2} 
                    \Bigg)^2 -1 
                \label{eq:Gladman_eh}
            \end{eqnarray}
    \end{enumerate}
These Newtonian conditions will be compared with our numerical results in \S\ref{sec:Result}.
In the following sections, we extend inequality~\eqref{eq:Sundman3} to include GR effects in the post-Newtonian approximation.
Finding a nice parameterization of the resultant inequality as for the Newtonian case 
is a big challenge and will be deferred to a future work.

\subsection{\label{subsec:Hill_PN} Post-Newtonian Analysis}

Now we present one of the main results of this paper.
The equations of motion in the first-order post-Newtonian (1PN) approximation are called the Einstein-Infeld-Hofmann equations \cite{EIH38}: 
    \begin{widetext}
    	\begin{eqnarray}
        	\frac{ \mathrm{d} \bm{v}_{k}}{\mathrm{d} t}
        			&=&-G\sum_{n\neq k} m_{n}\frac{\bm{x}_{k} - \bm{x}_{n}}{|\bm{x}_{k} - \bm{x}_{n}|^{3}}
        				 \Bigg[ 
        						1-4 \frac{G}{c^2}\sum_{n'\neq k} \frac{m_{n'}}{|\bm{x}_{k} - \bm{x}_{n'}|}
        	\nonumber \\
                          		&-&\frac{G}{c^2}\sum_{n'\neq n} \frac{m_{n'}}{|\bm{x}_{n} - \bm{x}_{n'}|} 
        						\left \{
                                     		   1-\frac{(\bm{x}_{k} - \bm{x}_{n}) \cdot (\bm{x}_{n} - \bm{x}_{n'})}
                                             {2|\bm{x}_{n} - \bm{x}_{n'}|^{2}} 
        						\right \} 
            \nonumber \\
        						&+&\left( \frac{|\bm{v}_{k}|}{c} \right)^{2} + 2\left( \frac{|\bm{v}_{n}|}{c} \right)^{2} 
        						-4 \frac{\bm{v}_{k} \cdot \bm{v}_{n}}{c^2}
                          		-\frac{3}{2} 
                          		\left\{ 
        								   \frac{(\bm{x}_{k} - \bm{x}_{n})}{|\bm{x}_{k} - \bm{x}_{n}|} \cdot \frac{\bm{v}_{n}}{c} 
        						\right\}^{2}   
                         \Bigg]  
            \nonumber \\
                         &-&\frac{G}{c^2} \sum_{n \neq k} \frac{m_{n}(\bm{v}_{k}-\bm{v}_{n})}{|\bm{x}_{k} - \bm{x}_{n}|^{3}}
                        		(\bm{x}_{k} - \bm{x}_{n}) \cdot (3\bm{v}_{n}-4\bm{v}_{k}) 
        	\nonumber \\
        				&-&\frac{7}{2} \frac{G^{2}}{c^2} \sum_{n \neq k}\frac{m_{n}}{|\bm{x}_{k} - \bm{x}_{n}|}
                        		\sum_{n'\neq n} \frac{m_{n'} (\bm{x}_{n} - \bm{x}_{n'})} {|\bm{x}_{n} - \bm{x}_{n'}|^{3}}\,. 
        	\label{eq:EIH}
         \end{eqnarray} 
    \end{widetext}
The 1PN Hamiltonian and linear momentum of a general $N$-body system are obtained from Eq.~\eqref{eq:EIH} as
    \begin{eqnarray}
        \mathcal{H}_\mathrm{PN} &=& \frac{1}{2} \sum_j m_j \Big( v_j^2 - \sum_{i\neq j}\frac{Gm_i}{r_{ij}} \Big) 
        \nonumber \\
            &+&\frac{1}{c^2}\sum_j m_j 
                    \Bigg[ 
                            \frac{3}{8}v_j^4 +\frac{3}{2} v_j^2 \sum_{i \neq j} \frac{Gm_i}{r_{ij}} 
        \nonumber \\
                            &+&\frac{1}{2}\sum_{i \neq j} \sum_{k \neq j} \frac{G^2m_im_k}{r_{ij}r_{jk}} 
                            -\frac{1}{4} \sum_{i \neq j} \frac{Gm_i}{r_{ij}} 
                                \{ 
                                    7\bm{v}_i \cdot \bm{v}_j
        \nonumber \\
                                    &+& ( \bm{v}_i \cdot \bm{n}_{ji})(\bm{v}_j \cdot \bm{n}_{ji}) 
                                \}
                    \Bigg] ,
        \label{eq:PNHamiltonian}
    \end{eqnarray}
    \begin{eqnarray}
        \bm{P}_j &=& m_j\bm{v}_j 
                    + \Bigg[ 
                            \frac{1}{2c^2} m_j\bm{v}_j \Big( v_j^2 - \sum_{i\neq j}\frac{Gm_i}{r_{ij}} \Big) 
        \nonumber \\
                            &-&\frac{G}{2c^2}\sum_{i\neq j}\frac{m_im_j}{r_{ij}} (\bm{v}_j \cdot \bm{n}_{ji}) \bm{n}_{ji}
                      \Bigg] ,
        \label{eq:PNmomentum}
    \end{eqnarray}
where the subscript $j$ means the $j$-th object and runs from 1 to 3 for the three-body system.
The total angular momentum $\bm{J}$ is defined as
    \begin{equation}
        \bm{J} = \sum_j \bm{r}_j \times \bm{P}_j .
        \label{eq:PNJ}
    \end{equation}
The total energy and total angular momentum given by Eqs.~\eqref{eq:PNHamiltonian} and \eqref{eq:PNJ}, respectively, are conserved quantities of the system.

The target in this paper is the triple systems that have a central object with a large mass $m_1$ like SMBH or IMBH and two orbiting objects with much smaller masses $m_2, m_3 \ll m_1$.
In this limited case, the barycenter of the system sits almost on the central object, and if the coordinate origin is set on the barycenter, the following approximate relations hold:
    \begin{eqnarray}
        && \bm{r}_1 \approx \bm{0}, \\
        && \bm{r}_2 \approx \bm{r}_2-\bm{r}_1, \\
        && \bm{r}_3 \approx \bm{r}_3-\bm{r}_1, \\
        && \bm{v}_1 \approx \bm{0}, \\
        && \bm{v}_2 \approx \bm{v}_2-\bm{v}_1, \\
        && \bm{v}_3 \approx \bm{v}_3-\bm{v}_1.
    \end{eqnarray}
With these approximations, the 3-body 1PN Hamiltonian Eq.~\eqref{eq:PNHamiltonian} can be recast into the following form: 
    \begin{equation}
        \mathcal{H}_\mathrm{PN} \approx m_2 \mathcal{H}_\mathrm{rel}^{(1-2)} +m_3 \mathcal{H}_\mathrm{rel}^{(1-3)} - \frac{Gm_2m_3}{r_{23}} ,
        \label{eq:PNH_divided}
    \end{equation}
where $\mathcal{H}_\mathrm{rel}^{(1-2)}$ and $\mathcal{H}_\mathrm{rel}^{(1-3)}$ are the 1PN specific Hamiltonians for the relative motions $\bm{r}_2-\bm{r}_1$ and $\bm{r}_3-\bm{r}_1$, respectively.
Here we neglected 1PN correction terms proportional to $m_2m_3$ or $m_1m_2m_3$ because they are formally of the order of $m_2(\sim m_3)/m_1 \times$ the dominant 1PN corrections, $m_2\mathcal{H}_\mathrm{rel}^{(1-2)}$ and $m_3\mathcal{H}_\mathrm{rel}^{(1-3)}$, and are confirmed by direct numerical evaluations to be always smaller by a factor of  $10^{6}$ in our simulations indeed.
Each 1PN specific Hamiltonian $\mathcal{H}_\mathrm{rel}$ as well as the specific linear momentum $\bm{p}$ of the relative motion $\bm{r} = \bm{r}_i-\bm{r}_1$ were derived by \citet{Richardson88} as
    \begin{eqnarray}
        \mathcal{H}_\mathrm{rel} &=& \frac{1}{2} \bm{p} \cdot \bm{p} - \frac{G(m_1+m_i)}{r} 
                                        -\frac{1}{c^2}  
                                            \Big[
                                                \sigma_0 (\bm{p} \cdot \bm{p})^2
        \nonumber \\
                                 &&             + \frac{\sigma_1}{r} \bm{p} \cdot \bm{p}
                                                + \frac{\sigma_2}{r^2} + \frac{\sigma_3}{r^3} ( \bm{r} \cdot \bm{p} ) ^2
                                            \Big] ,
        \label{eq:relativeH}
    \end{eqnarray}
    \begin{equation}
        \bm{p} = \bm{v} + \frac{1}{c^2} \Big[
                                            4\sigma_0 v^2 \bm{v}+ \frac{2\sigma_1}{r} \bm{v} 
                                            + \frac{2\sigma_3}{r^3} (\bm{r} \cdot \bm{v}) \bm{r}
                                        \Big].
        \label{eq:relativep}
    \end{equation}
The coefficients in Eqs.~\eqref{eq:relativeH} and \eqref{eq:relativep} are given as 
    \begin{eqnarray}
        && \sigma_0 = \frac{1-3\chi}{8}, \\
        && \sigma_1 = \frac{G(m_1+m_i)(3+\chi)}{2}, \\
        && \sigma_2 = -\frac{G^2 (m_1+m_i)^2}{2}, \\
        && \sigma_3 = \frac{G(m_1+m_i)\chi}{2}, \\
        && \chi = \frac{m_1m_i}{(m_1+m_i)^2}.
    \end{eqnarray}
In Eq.~\eqref{eq:relativeH} we may make the following replacements $G(m_1+m_2) \approx Gm_1$, $\bm{r}_2-\bm{r}_1 \approx \bm{r}_2$ and $\bm{v}_2-\bm{v}_1 \approx \bm{v}_2$. 
Denoting the terms of the order of $1/c^2$ in $\mathcal{H}_\mathrm{rel}$ as $X(\bm{r}, \bm{p})$, we write $\mathcal{H}_\mathrm{rel}^{(1-2)}$ and $\mathcal{H}_\mathrm{rel}^{(1-3)}$ as 
    \begin{equation}
        \mathcal{H}_\mathrm{rel}^{(1-2)} \approx \frac{1}{2} \bm{p}_2 \cdot \bm{p}_2 - \frac{Gm_1}{r_{12}} + X^{(1-2)}(\bm{r}_2, \bm{p}_2),
        \label{eq:relativeH_2}
    \end{equation}
        \begin{equation}
        \mathcal{H}_\mathrm{rel}^{(1-3)} \approx \frac{1}{2} \bm{p}_3 \cdot \bm{p}_3 - \frac{Gm_1}{r_{13}} + X^{(1-3)}(\bm{r}_3, \bm{p}_3).
        \label{eq:relativeH_3}
    \end{equation}
Substituting these expressions in Eq.~\eqref{eq:PNH_divided}, we can rewrite the total Hamiltonian as
    \begin{eqnarray}
        \mathcal{H}_\mathrm{PN} &\approx& \frac{1}{2} m_2 p_2^2 + \frac{1}{2} m_3 p_3^2 + U 
        \nonumber \\
                                    &+& m_2 X^{(1-2)}(\bm{r}_2, \bm{p}_2) + m_3 X^{(1-3)}(\bm{r}_3, \bm{p}_3),
    \end{eqnarray}
where $U$ is the Newtonian gravitational potential Eq.\eqref{eq:N_potential}.

As remarked in \S\ref{subsec:Hill_N}, the two vectors in Sundman's inequality \eqref{eq:Sundman} can be chosen arbitrarily.
In the post-Newtonian analysis, we take the specific linear momentum $\bm{p}$ instead of the velocity $\bm{v}$ to obtain
    \begin{eqnarray}
        && \left( \sum_j m_j r_j^2 \right) \left(  \sum_j m_j p_j^2 \right) 
        \nonumber \\
        && \ge  \left| \sum_j m_j \bm{r}_j \times \bm{p}_j \right|^2 + \left| \sum_j m_j \bm{r}_j \cdot \bm{p}_j \right|^2.
        \label{eq:Sundman_1PN}
    \end{eqnarray}
Note that the terms with $j=1$ are almost vanishing for the systems of our concern in this paper.
Following the procedure in \S\ref{subsec:Hill_N}, we can further rewrite inequality~\eqref{eq:Sundman_1PN} in terms of the characteristic quantities of the system as
        \begin{eqnarray}
            &2&\left( \sum_j m_j r_j^2 \right) 
                \Big(  
                        \mathcal{H}_\mathrm{PN} - U - m_2 X^{(1-2)}(\bm{r}_2, \bm{p}_2) 
            \nonumber \\
            &&          - m_3 X^{(1-3)}(\bm{r}_3, \bm{p}_3) 
                \Big) 
                        \ge J^2 .
            \label{eq:Sundman_1PN_3}
        \end{eqnarray}

The above inequality is the relativistic counterpart of inequality~\eqref{eq:Sundman3}.
However, this is not very convenient.
The difficulty here is that $X^{(1-2)}(\bm{r}_2, \bm{p}_2)$ and $X^{(1-3)}(\bm{r}_3, \bm{p}_3)$ depend not only on the position but also on the specific linear momentum.
In order to obtain the allowed or forbidden region, 
they need to be approximated somehow with the functions of the position alone.
Here we propose to apply the virial theorem in the 1PN approximation, which was derived by \citet{Chandrasekhar608},
individually to the two-body systems consisting of the central object and one of the orbiting objects:
    \begin{equation}
        v_i^2 \approx \frac{Gm_1}{r_{i}} \left( 1- \frac{3}{c^2} \frac{Gm_1}{r_{i}} \right)
        \label{eq:virial}
    \end{equation}
In fact, the last term in Eq.~\eqref{eq:virial} can be neglected because 
it is employed in those terms that are already of the 1PN order and, as a result, becomes of
higher PN orders.
When the higher order term is neglected, this approximation is reduced to the Newtonian virial relation. 
Rigorously speaking, this relation holds only for the average over the orbital cycle unless the orbit is circular.
In this paper, however, we use this relation even for elliptic orbits pointwise as an approximation and it turns out it is very successful.

Substituting this approximation in Eq.~\eqref{eq:relativep} and employing the result in the definition of $X^{(1-i)}(\bm{r}_i, \bm{p}_i)$, we obtain
    \begin{equation}
        X^{(1-i)}(\bm{r}_i, \bm{p}_i) \approx -\frac{9}{8} \frac{1}{c^2} \frac{G^2m_1^2}{r_i^2} + \mathcal{O} \left( \frac{v_i^4}{c^4} \right) .
    \end{equation}
Note that the angle $\theta$ between $\bm{r}_i$ and $\bm{p}_i$, 
always appears as $p^2(3 +\chi \cos^2 \theta)$ in $X^{(1-i)}(\bm{r}_i, \bm{p}_i)$ and 
is neglected because $\chi \ll 1$. 
With this approximated $X^{(1-i)}(\bm{r}_i, \bm{p}_i)$, 
Eq.~\eqref{eq:Sundman_1PN_3} is also approximately written as
        \begin{eqnarray}
            2\left( \sum_j m_j r_j^2 \right) 
                \Bigg\{  
                        \mathcal{H}_\mathrm{PN} - U 
                     + \frac{9}{8} \frac{G^2m_1^2}{c^2} && 
                        \left( 
                                \frac{m_2}{r_2^2} +  \frac{m_3}{r_3^2} 
                        \right) 
                \Bigg\} 
                 \nonumber \\
            &&
                        \ge J^2 .
            \label{eq:Sundman_1PN_virial}
        \end{eqnarray}
This inequality is more like the Newtonian counterpart, Eq.~\eqref{eq:Sundman3}, and is the basis for the following analysis.

Next we bound
$\mathcal{H}_\mathrm{PN}$ from above and 
$J^2$ from below in inequality~\eqref{eq:Sundman_1PN_virial}, 
employing the initial orbital elements as
    \begin{eqnarray}
        \mathcal{H}_\mathrm{PN} \lesssim &-& \frac{Gm_1m_2}{2a_\mathrm{in}} -\frac{Gm_1m_3}{2a_\mathrm{out}}
        \nonumber \\
                                     &+& \frac{19}{8} \frac{G^2m_1^2}{c^2} 
                                            \Big\{ 
                                                \frac{m_2}{a_\mathrm{in}^2(1-e_\mathrm{in})^2} 
        \nonumber \\
                                     &&          + \frac{m_3}{a_\mathrm{out}^2(1-e_\mathrm{out})^2} 
                                            \Big\} ,
        \label{eq:H_1PN_approx}
    \end{eqnarray}
    \begin{eqnarray}
        J^2 \gtrsim &J_\mathrm{in,N}^2& \left[ 1 + \frac{7}{c^2} \frac{Gm_1}{a_\mathrm{in}(1+e_\mathrm{in})} \right]
                 + 2 J_\mathrm{in,N} J_\mathrm{out,N} 
        \nonumber \\
                    &\times&
                        \Bigg[ 
                                1 + \frac{7}{2} \frac{Gm_1}{c^2}
                                \left( \frac{1}{a_\mathrm{in}(1+e_\mathrm{in})} + \frac{1}{a_\mathrm{out}(1+e_\mathrm{out})} \right)
                        \Bigg]
        \nonumber \\
                 &+& J_\mathrm{out,N}^2 \left[ 1 + \frac{7}{c^2} \frac{Gm_1}{a_\mathrm{out}(1+e_\mathrm{out})} \right],
        \label{eq:J_1PN_approx}
    \end{eqnarray}
where $J_\mathrm{in,N}$ and $J_\mathrm{out,N}$ are given in Eqs.~\eqref{eq:J_in} and \eqref{eq:J_out}.
These are corresponding to Eqs.~\eqref{eq:H_N} and \eqref{eq:J^2_N}, respectively.
The detailed derivations of these estimations are presented as follows.

By using the virial relation Eq.~\eqref{eq:virial}, the Hamiltonian $\mathcal{H}_\mathrm{PN}$ is approximately written as
    \begin{eqnarray}
        \mathcal{H}_\mathrm{PN} &\approx& \frac{1}{2} m_2 v_2^2 + \frac{1}{2} m_3 v_3^2 + U 
        \nonumber \\
                                    &+& \frac{19}{8} \frac{1}{c^2} 
                                        \left( 
                                            \frac{G^2m_1^2m_2^2}{r_2^2} + \frac{G^2m_1^2m_3^2}{r_3^2} 
                                        \right).
        \label{eq:H_1PN_eval1}
    \end{eqnarray}
The Newtonian orbital energy can be rewritten with the initial semi-major axes $a_\mathrm{in}$ and $a_\mathrm{out}$ as
    \begin{eqnarray}
        \frac{1}{2}m_2v_2^2-\frac{Gm_1m_2}{r_2} &\approx& \frac{1}{2}\frac{m_1m_2}{m_1+m_2}v_{12}^2-\frac{Gm_1m_2}{r_{12}}
        \nonumber \\
                                                &=& -\frac{Gm_1m_2}{2a_\mathrm{in}},
    \end{eqnarray}
    \begin{eqnarray}
        \frac{1}{2}m_3v_2^3-\frac{Gm_1m_3}{r_3} &\approx& \frac{1}{2}\frac{m_1m_3}{m_1+m_3}v_{13}^2-\frac{Gm_1m_3}{r_{13}}
        \nonumber \\
                                                &=& -\frac{Gm_1m_3}{2a_\mathrm{out}}.
    \end{eqnarray}
The term for the gravitational interaction between $m_2$ and $m_3$, $Gm_2m_3/r_{23}$, can be neglected because it is much smaller than $Gm_1m_2/r_{12}$ and $Gm_1m_3/r_{13}$.
In order to use inequality \eqref{eq:Sundman_1PN_virial} we should bound the Hamiltonian from above.
The last term in Eq.~\eqref{eq:H_1PN_eval1} can be evaluated with following relation
    \begin{equation}
        \frac{1}{r^2} \leq \frac{1}{a^2(1-e)^2}.
        \label{eq:app_r1}
    \end{equation}
In this evaluation, we used the periapsis distance in the Kepler orbit as the minimum value of the distance between the central and orbiting objects.
The Hamiltonian is now estimated as Eq.~\eqref{eq:H_1PN_approx}.

The square of magnitude of the angular momentum $J^2$ can be estimated similarly.
The total angular momentum $\bm{J}$ is written as
    \begin{eqnarray}
        \bm{J} &=& \bm{J}_\mathrm{in,N} 
                    \left[ 
                            1 + \frac{1}{c^2} 
                            \left(
                                    \frac{v_2^2}{2} + \frac{3Gm_1}{r_2}
                            \right) 
                    \right]
        \nonumber \\
               &+& \bm{J}_\mathrm{out,N} 
                    \left[ 
                            1 + \frac{1}{c^2} 
                            \left(
                                    \frac{v_3^2}{2} + \frac{3Gm_1}{r_3}
                            \right) 
                    \right],
    \end{eqnarray}
where $\bm{J}_\mathrm{in,N} = m_2 \bm{r}_2 \times \bm{v}_2$ and $\bm{J}_\mathrm{out,N} = m_3 \bm{r}_3 \times \bm{v}_3$ are the Newtonian angular momenta of the inner and outer orbits, whose magnitudes are expressed with the orbital elements in Eqs.~\eqref{eq:J_in} and \eqref{eq:J_out}.
The magnitude of the total angular momentum squared $J^2$ is given as
    \begin{eqnarray}
        J^2 &=& J_\mathrm{in,N}^2 
                    \left[ 
                            1 + \frac{2}{c^2} 
                            \left(
                                    \frac{v_2^2}{2} + \frac{3Gm_1}{r_2}
                            \right) 
                    \right]
        \nonumber \\
               &+& 2 \bm{J}_\mathrm{in,N} \cdot \bm{J}_\mathrm{out,N} 
                    \Bigg[ 
                            1 + \frac{1}{c^2} 
                            \Big\{
                                    \frac{v_2^2 + v_3^2}{2} 
        \nonumber \\
                                    &+& 3Gm_1 \left( \frac{1}{r_2} + \frac{1}{r_3} \right)
                            \Big\} 
                    \Bigg]
        \nonumber \\
               &+& J_\mathrm{out,N} 
                    \left[ 
                            1 + \frac{2}{c^2} 
                            \left(
                                    \frac{v_3^2}{2} + \frac{3Gm_1}{r_3}
                            \right) 
                    \right]
        \nonumber \\
               &+& \mathcal{O}\left(  \frac{v^4}{c^4} \right) .
        \label{eq:J_1PN_eval1}
    \end{eqnarray}
The scalar product of the inner and outer angular momenta can be replaced as $\bm{J}_\mathrm{in,N} \cdot \bm{J}_\mathrm{out,N} = J_\mathrm{in,N} J_\mathrm{out,N}$ because the systems considered in this paper have the coplanar prograde orbits and the two angular momenta are aligned with each other.
We employ the virial relation Eq.~\eqref{eq:virial} again in Eq.~\eqref{eq:J_1PN_eval1} as
    \begin{eqnarray}
                J^2 &\approx& J_\mathrm{in,N}^2 
                    \left[ 
                            1 + 7\frac{1}{c^2} \frac{Gm_1}{r_2}
                    \right]
        \nonumber \\
               &+& 2 J_\mathrm{in,N} J_\mathrm{out,N} 
                    \left[ 
                            1 + \frac{7}{2} \frac{Gm_1}{c^2} \left( \frac{1}{r_2} +\frac{1}{r_3} \right)
                    \right]
        \nonumber \\
               &+& J_\mathrm{out,N} 
                    \left[ 
                            1 + 7\frac{1}{c^2} \frac{Gm_1}{r_3}
                    \right] 
                    + \mathcal{O}\left(  \frac{v^4}{c^4} \right) .
        \label{eq:J_1PN_eval2}
    \end{eqnarray}
We should bound $J^2$ from below this time to use the result in inequality \eqref{eq:Sundman_1PN_virial}. 
The following relation is employed:
    \begin{equation}
        \frac{1}{r} \geq \frac{1}{a(1+e)}.
        \label{eq:app_r2}
    \end{equation}
Here the maximum value of the distance between the central and orbiting objects is set as the apoapsis distance in the Kepler orbit.
Then the total angular momentum squared is estimated as Eq.~\eqref{eq:J_1PN_approx}.

Employing these inequalities~\eqref{eq:H_1PN_approx} and \eqref{eq:J_1PN_approx} in inequality~\eqref{eq:Sundman_1PN_virial} and fixing the positions of two objects, we finally obtain the allowed and forbidden regions of the remaining body.
As explained earlier for the Newtonian case, the existence of the forbidden region between two orbits may be interpreted as a sufficient condition of the 1PN Hill stability for the triple system with a massive central object and two orbiting objects with much smaller masses.
We evaluate inequalities~\eqref{eq:Sundman_1PN_virial}, \eqref{eq:H_1PN_approx} and  \eqref{eq:J_1PN_approx} numerically to obtain the forbidden region for some models in \S\ref{sec:Result}.
We also compare the sufficient condition so obtained with the results of numerical three-body simulations in the 1PN approximation to validate our criterion.
   
\section{\label{sec:Num}Numerical Simulations}
In order to test the relativistic Hill stability condition we obtained, we conduct some numerical simulations following \citet{Chambers96}, who computed Newtonian
orbital evolutions of multi-body systems of various initial orbital separations until instability occurs in the sense of Hill stability.
The onset of the instability was judged from the orbital separation during the simulation.
From the relation between the initial orbital separation and the time when the system becomes unstable we can obtain the condition of the Hill Stability.

In the following, we conduct similar simulations both in the Newtonian and the first-order post-Newtonian approximations and compare the results.
Note that neglected higher order terms in the PN approximation may have some important effects on the Hill stability.
For example, some authors recently have studied the 1.5 PN order effects, that is, the spin-orbit coupling called the Lense-Thrring effect, on the orbital evolution of the hierarchical triple systems \cite{Fang19a, Fang19b, Liu19}.
The 2.5 PN order effects corresponding to the GW emission may be also important:
it extracts energy from the inner orbit more efficiently than from the outer orbit \cite{Peters63} and, as a result, the orbital separation will become larger, thus affecting the Hill stability.
In this paper, however, we ignore these interesting higher order effects and focus on the 1PN effect as a first step.
We will give a rough estimation of these effects in \S\ref{sec:Result}, though.
They will be investigated in detail in future works.

Our numerical models of relativistic three-body systems are divided into two groups: 
those with an SMBH (we call it the SMBH group) as a central object and the others with an IMBH (we refer to it as the IMBH group). 
Each group has three models: 
circular, small-eccentricity (small-$e$) and large-eccentricity (large-$e$) models according to the classification by \citet{Gladman93}; 
in the circular model, the inner and outer orbits are both circular, whereas in the low- and high-eccentricity models, the two orbits have eccentricities that satisfy $e<\mu^{1/3}$ and $e>\mu^{1/3}$, respectively.

The important parameters in the initial conditions are summarized for all the models in Table \ref{tab:param}.
    \begin{table*}
        \caption{\label{tab:param}
                The important parameters in the initial conditions for all models treated in this paper.
                The third and fourth columns, $m_1$ and $m_i$, are the masses of the central object and the orbiting objects in the three-body system. 
                The subscript $i$ runs from 2 to 3.
                In this paper, we set $m_2 = m_3$.
                The fifth column, $a_\mathrm{in}$, is the semi-major axis of the inner orbit.
                The semi-major axis of the outer orbit, $a_\mathrm{out}$, is determined 
                from the parameter $\Delta$ as explained in the text.
                The sixth column, $e$, is the eccentricity, which is assumed to be common to the inner and outer orbits.
                The last column, $\omega$, is the argument of periapsis of the two orbits, which are assumed to be the same.
                Note that in the circular orbit, we cannot define the argument of periapsis.
                The information about the other orbital elements, for example, the inclinations and the mean anomalies are given in the text.
                }
        \begin{ruledtabular}
		    \begin{tabular}{ccccccc}
				 group &    model   & $m_1[M_\odot]$ & $m_i[M_\odot]$  &$a_\mathrm{in}$[au] &    $e$   & $\omega$[deg]\\
				\hline
				 SMBH  &   circular &     $10^6$     &     $1.0$       &        $1.0$       &   0      &     -       \\
                 SMBH  &  small-$e$ &     $10^6$     &     $1.0$       &        $1.0$       &   0.009  &     0       \\ 
                 SMBH  &  large-$e$ &     $10^6$     &     $1.0$       &        $1.0$       &   0.1    &     0       \\
                 IMBH  &   circular &     $10^3$     &     $1.0$       &        $0.1$       &   0      &     -       \\
                 IMBH  &  small-$e$ &     $10^3$     &     $1.0$       &        $0.1$       &   0.009  &     0       \\ 
                 IMBH  &  large-$e$ &     $10^3$     &     $1.0$       &        $0.1$       &   0.2    &     0       \\
		    \end{tabular}
		\end{ruledtabular}
	\end{table*}
There are six orbital elements for each orbit in general.
We use the so-called Kepler elements: 
the semi-major axis $a$, the eccentricity $e$, the inclination $i$, the argument of periastron $\omega$, the longitude of ascending node $\Omega$, and the mean anomaly $M$.
In the SMBH group, we fix the inner semi-major axis $a_\mathrm{in}$ to 1.0 au while in the IMBH group, $a_\mathrm{in}$ is determined so that the period of the inner orbit should be the same as the counterpart in the SMBH group to facilitate comparison.
All the models have coplanar and prograde orbits, that is, the relative inclination between the inner and outer orbits is zero.
The longitude of the ascending node $\Omega$ cannot be defined in this case.

Note that the Newtonian Hill stability in non-coplanar systems are investigated in detail by \citet{Grishin17}.
They showed that for highly inclined hierarchical three-body systems, the Kozai-Lidov mechanism operates and affects the stability.
It is known, on the other hand, that GR suppresses the Kozai-Lidov mechanism in some parameter regimes \cite{Blaes02, Anderson17}.
Although its ramification for the stability is an interesting issue, it is beyond the scope of our paper and will be addressed in future.

As mentioned repeatedly, we are concerned in this paper with the relation between the onset time of the orbital instability and the initial orbital separation $\Delta$, which is defined as the difference of 
the semi-major axes in the units of the mutual Hill radius $R'_{\mathrm{Hill}}$:
    \begin{equation}
        a_\mathrm{out}-a_\mathrm{in}=\Delta R'_{\mathrm{Hill}},    
        \label{Eq:separation}
    \end{equation}
where $R'_{\mathrm{Hill}}$ is defined as
    \begin{equation}
        R'_{\mathrm{Hill}} \equiv \left( \frac{ \mu_2 + \mu_3 }{3} \right)^{\frac{1}{3}} \frac{a_\mathrm{in} + a_\mathrm{out}}{2}    
        \label{mutual Hill}
    \end{equation}
with $\mu_i$ being the ratio of the mass of the $i$-th orbiting object to the mass of the central object. 
For each model, we change the value of $\Delta$ from 1.0 by an increment of 0.1 and compute the orbital evolution both in the Newtonian and 1PN approximations.
The initial mean anomalies of the two orbiting objects $M_\mathrm{in}$ and $M_\mathrm{out}$ are set randomly except that they should be separated by  at least $20^{\circ}$.
For each value of $\Delta$ we perform three runs with different combinations of mean anomalies.

The Kepler elements are transformed to the positions and velocities in the Cartesian coordinates of the constituent bodies, the detail of which is given in Appendix \ref{app:initial} (see also, e.g., \citet{SSD2000}).
The Newtonian and 1PN (Eq.~\eqref{eq:EIH}) equations of motion are numerically integrated by using the 6-th order implicit Runge-Kutta (IRK) method \cite{Butcher64}
\footnote{There are different methods to calculate long-term relativistic orbital evolutions.
One of the most commonly used formulae is the so-called double-averaging method, which is derived by averaging the Hamiltonian with respect to both the inner and outer orbits.
In this method, the Newtonian gravitational interaction between the inner and outer orbits is normally treated perturbatively and \citet{Will14a, Will14b} stressed the importance of the ‘cross terms’ between the above Newtonian terms and the post-Newtonian terms, which are commonly omitted. 
We remark that, in our calculations, we integrate the equations of motion directly and the effects of the cross terms are automatically included.}. 
Each run is continued up to either the onset of instability or $10^6$ yrs.
When the integration is completed, we reconvert the positions and velocities at each timestep into the orbital elements of the {\it osculating orbit}, the detail of which is explained in Appendix \ref{app:post}. 

We decide that the instability sets in when the difference of the distance of the periastron of the outer orbit and that of apoastron of the inner orbit becomes smaller than one of the Hill radii of the two orbiting objects: 
    \begin{equation}
        a_\mathrm{out}(1-e_\mathrm{out}) - a_\mathrm{in}(1+e_\mathrm{in}) < R_{\mathrm{Hill},i},    
        \label{eq:judge}
    \end{equation} 
where $R_{\mathrm{Hill},i}$ is the Hill radius of the $i$-th object defined as
    \begin{equation}
        R_{\mathrm{Hill},2} \equiv \left( \frac{\mu_{2}}{3} \right)^{\frac{1}{3}} a_\mathrm{in} 
        \label{eq:Hill}
    \end{equation}
for the second object and is given similarly for the third object with $\mu_2$ and $a_\mathrm{in}$ being replaced with $\mu_3$ and $a_\mathrm{out}$, respectively.
We remark that some authors employed a different criterion of close encounter: the separation of two orbits should become smaller than the mutual Hill radius.
The Hill radius and the mutual Hill radius are not much different from each other, however. 
We hence do not think that the change of the criterion would produce qualitatively different results.
As mentioned earlier, if Eq.~\eqref{eq:judge} is satisfied at some point in the simulation, we record the time as the onset time of instability $T_\mathrm{stab}$.
If, on the other hand, the system has a stable evolution up to $10^6$ yrs in all the three calculations for the same $\Delta$ but different initial mean anomalies for the consecutive three values of $\Delta$, we stop the calculation for that model.

\section{\label{sec:Result}Result \& Discussion}
\subsection{\label{subsec:res_SMBH} SMBH group}

    \begin{figure}
        \centering
        \includegraphics[width=8.5cm,clip]{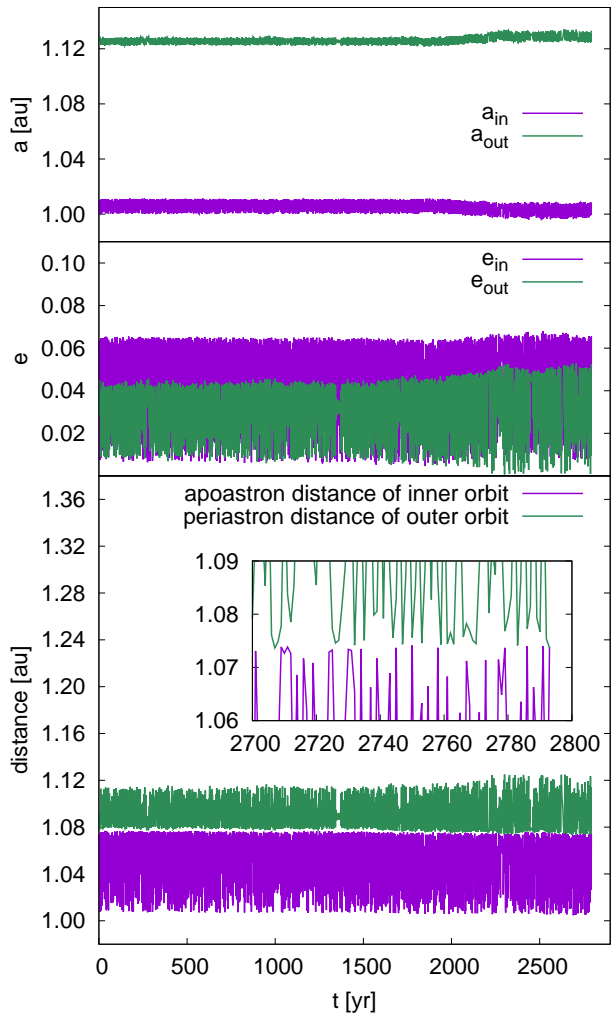}
        \caption{
                The 1PN-evolution of the orbital elements for the SMBH small-$e$ model with $\Delta=13.3$.
                The top and middle panels show the evolutions of the semi-major axes and eccentricities, respectively.
                The bottom panel exhibits the time variations of the apoastron distance of the inner orbit and the periastron distance of the outer orbit.
                The inset is the zoom-in to the onset time of the instability.
                The purple and green lines represent the inner and outer orbital elements, respectively.
                The onset time of instability $T_\mathrm{stab}$ is 2793 yrs in this case. 
                }
        \label{fig:orbit_ev}
    \end{figure}
We show the 1PN-evolutions of orbital elements of the SMBH small-$e$ model with $\Delta=13.3$ in Fig.~\ref{fig:orbit_ev} as an example of our simulations.
The evolutions of the semi-major axes and eccentricities are exhibited in the top and middle panels, respectively, whereas the evolutions of the apoastron distance of the inner orbit and of the periastron distance of the outer orbit are presented in the bottom panel.
We can see that both the semi-major axes and the eccentricities are fluctuating around their initial values until the separation between the apoastron distance of the inner orbit and the periastron distance of the outer orbit ceases to satisfy the criterion of the Hill stability.
In this case, $T_\mathrm{stab}$ is 2793 yrs. 

    \begin{figure}
        \centering
        \includegraphics[width=8.5cm,clip]{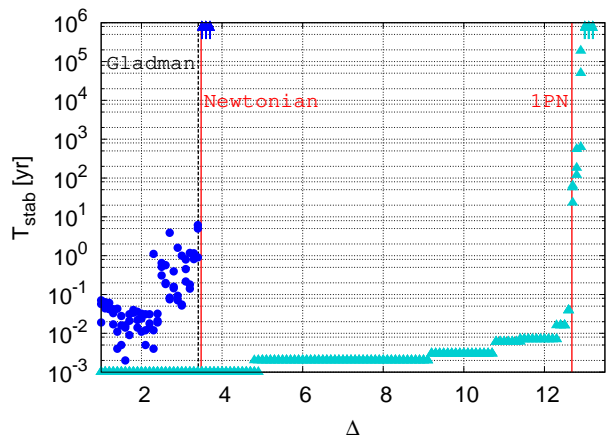}
        \caption{
                The relation between $\Delta$ and the onset time of instability for the circular model in the SMBH group.
                The blue dots show the results of the Newtonian calculations whereas the cyan triangles are the results from the 1PN calculations.
                See the text for details on other lines.
                }
        \label{fig:SMBH_circular_T}
    \end{figure}
    \begin{figure}
        \centering
        \includegraphics[width=8.5cm,clip]{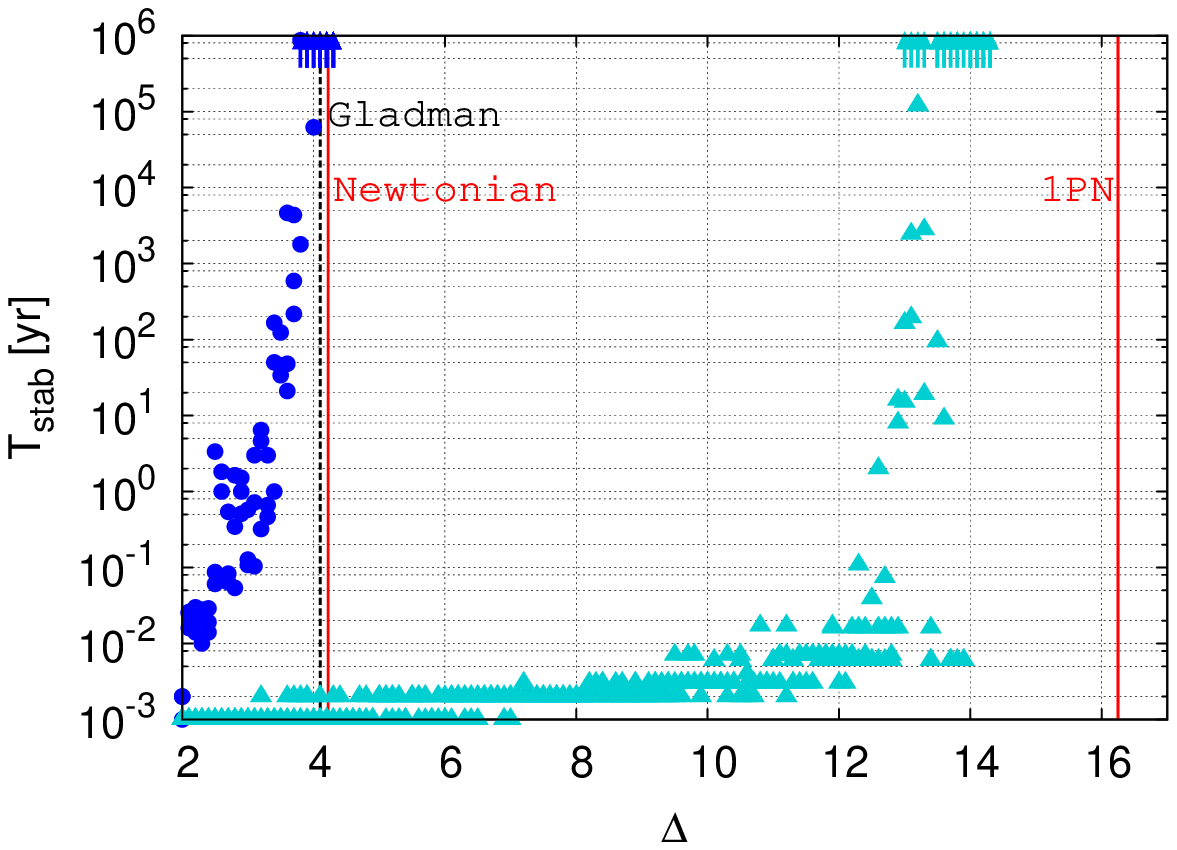}
        \caption{
                The same as Fig.~\ref{fig:SMBH_circular_T} but for the small-$e$ model in the SMBH group.
                }
        \label{fig:SMBH_e_small_T}
    \end{figure}
    \begin{figure}
        \centering
        \includegraphics[width=8.5cm,clip]{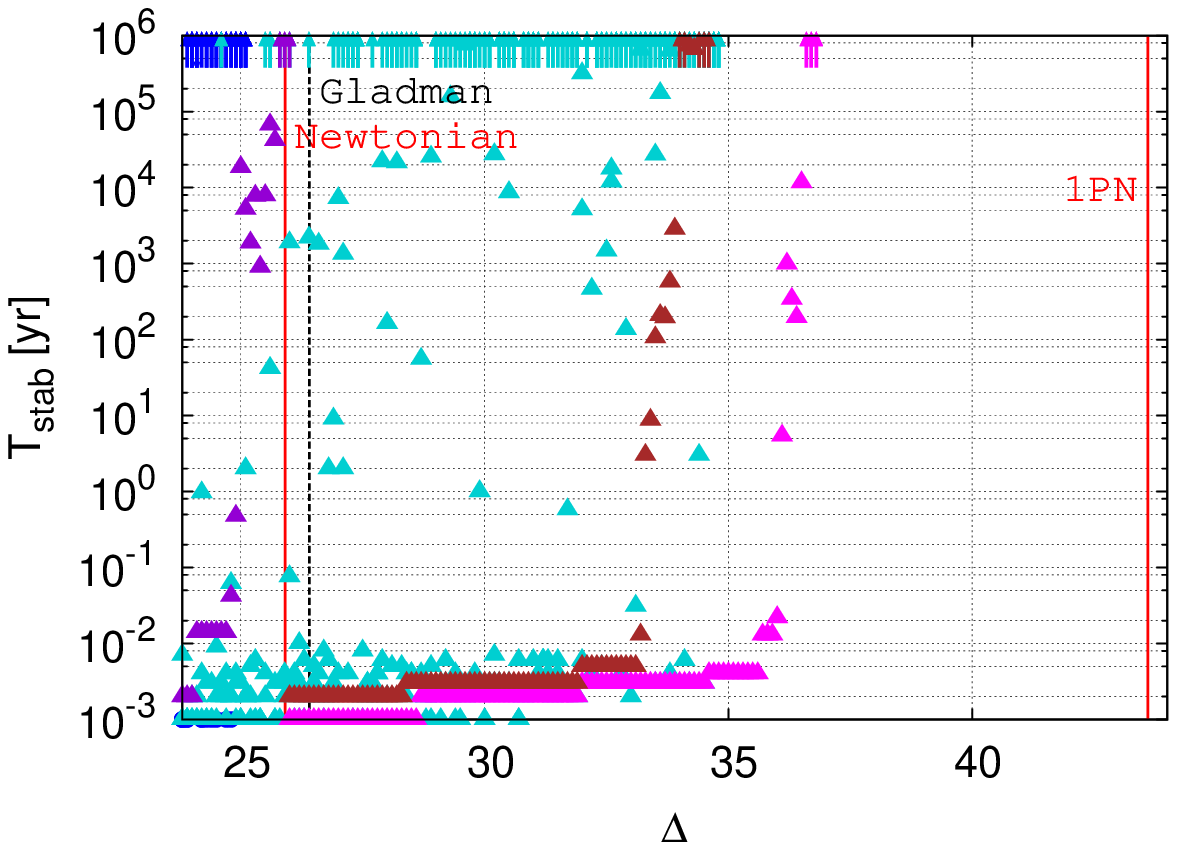}
        \caption{
                The same as Figs.~\ref{fig:SMBH_circular_T} and \ref{fig:SMBH_e_small_T} but for the large-$e$ model in the SMBH group.
                Three 1PN calculations with fixed initial mean anomalies are shown in the triangles with different colors.
                Magenta, brown and purple triangles are the results of the 1PN calculations with 
                $(M_\mathrm{in}, M_\mathrm{out})=(0^\circ, 0^\circ)$, $(0^\circ, 180^\circ)$ and $(180^\circ, 90^\circ)$, respectively.
                }
        \label{fig:SMBH_e_large_T}
    \end{figure}
The relations between $\Delta$ and $T_\mathrm{stab}$ are summarized in Figs.~\ref{fig:SMBH_circular_T} to \ref{fig:SMBH_e_large_T} for the models in the SMBH group, which correspond to the circular, small-$e$ and large-$e$ models, respectively.
In these figures, the blue dots show the Newtonian results while the cyan triangles are the results obtained by the 1PN calculations.
In all the figures, $T_\mathrm{stab}$ is shorter for the 1PN calculations than for the Newtonian ones, that is, the 1PN evolutions are more unstable than the Newtonian counterparts in the sense of Hill stability.

The periastron shift, which provides extra perturbations in the former, may be the cause of the earlier instability.
In fact, the timescale of the periastron shift $t_\mathrm{P}$ can be estimated from the Lagrange planetary equations (see e.g. \cite{SSD2000, Suzuki20}).
It is written with the 1PN averaged potential $V_\mathrm{1PN}=3G^2m_1^2/(c^2a_\mathrm{in}^2\sqrt{1-e_\mathrm{in}^2})$ (see e.g. \cite{Migaszewski11}) and the mean motion of the inner orbit $n_\mathrm{in}=\sqrt{Gm_1/a_\mathrm{in}^3}$ as
    \begin{eqnarray}
        t_\mathrm{P} &=& \frac{n_\mathrm{in}a_\mathrm{in}^2}{V_\mathrm{1PN}} 
                            \approx 
                            \frac{1}{3\pi} \frac{a_\mathrm{in}}{r_{g,1}}   
                              P_\mathrm{in}
                            \nonumber \\
                        &\sim& 2 \mathrm{day}
                                \left( \frac{a_\mathrm{in}}{1.0\mathrm{au}} \right)  
                                \left( \frac{r_{g,1}}{0.020 \mathrm{au}} \right)^{-1}
                                \left( \frac{P_\mathrm{in}}{0.365\mathrm{day}} \right),
                        \label{eq:t_PN}
    \end{eqnarray}
where $r_{g,1}=2Gm_1/c^2$ is the Schwarzschild radius of the central object and $P_\mathrm{in}$ is the period of the inner orbit.
We find that this timescale is not much longer than the period of the inner orbit, and, 
more importantly, it is smaller than $T_\mathrm{stab}$ in the Newtonian case as should be evident, e.g., for $\Delta<4.0$ in Fig.~\ref{fig:SMBH_circular_T}.
This suggests that the periastron shift affects indeed the 1PN Hill stability.

The time to the onset of instability $T_\mathrm{stab}$ grows almost monotonically in the circular and small-$e$ models whereas in the large-$e$ model, its behavior is more complicated.
This is because the initial mean anomalies become an important factor for the orbits with large eccentricities.
In Fig.~\ref{fig:SMBH_e_large_T}, we can confirm this by comparing the results of the 1PN calculations with initial mean anomalies fixed to three different values:
magenta, brown and purple triangles are the 1PN results for $(M_\mathrm{in}, M_\mathrm{out})=(0^\circ, 0^\circ)$, $(0^\circ, 180^\circ)$ and $(180^\circ, 90^\circ)$, respectively.
One observes that $T_\mathrm{stab}$ grows almost monotonically with the initial separation $\Delta$ when the initial mean anomaly is fixed.
For the models with $(M_\mathrm{in}, M_\mathrm{out})=(180^\circ, 90^\circ)$, $T_\mathrm{stab}$ grows rapidly around $\Delta=25.0$ 
whereas it
keeps small value until $\Delta = 35.0$ for $(M_\mathrm{in}, M_\mathrm{out})=(0^\circ, 0^\circ)$.
These two results 
are probably the extremes and encompasses the results with other mean anomalies.

The black dashed lines in these figures correspond to the sufficient conditions given by \citet{Gladman93} (Eqs.~\eqref{eq:Gladman_c} to \eqref{eq:Gladman_eh}).
As seen in the figures, Gladman's sufficient conditions are consistent with our Newtonian results
whereas they are clearly inconsistent with the 1PN results.
It is hence inappropriate to apply Gladman's Newtonian sufficient conditions for Hill stability to such compact multi-body systems containing a SMBH as considered here.

On the other hand, our new sufficient conditions for the Hill stability works much better as shown with red solid lines in these figures. 
They are excellent particularly for the circular orbits (see Fig.~\ref{fig:SMBH_circular_T}). 
In the case of the eccentric orbits, they tend to overestimate $\Delta$ somewhat (Figs.~\ref{fig:SMBH_e_small_T} and \ref{fig:SMBH_e_large_T}). 
Considering that the criterion is supposed to be a sufficient condition for the Hill stability and that the stability is rather sensitive to the initial mean anomaly as just mentioned, we think that our criterion is a substantial improvement from the Gladman's. 
We will return to these results later.

    \begin{figure*}
        \centering
        \begin{minipage}{8.5cm}
            \includegraphics[width=8.5cm]{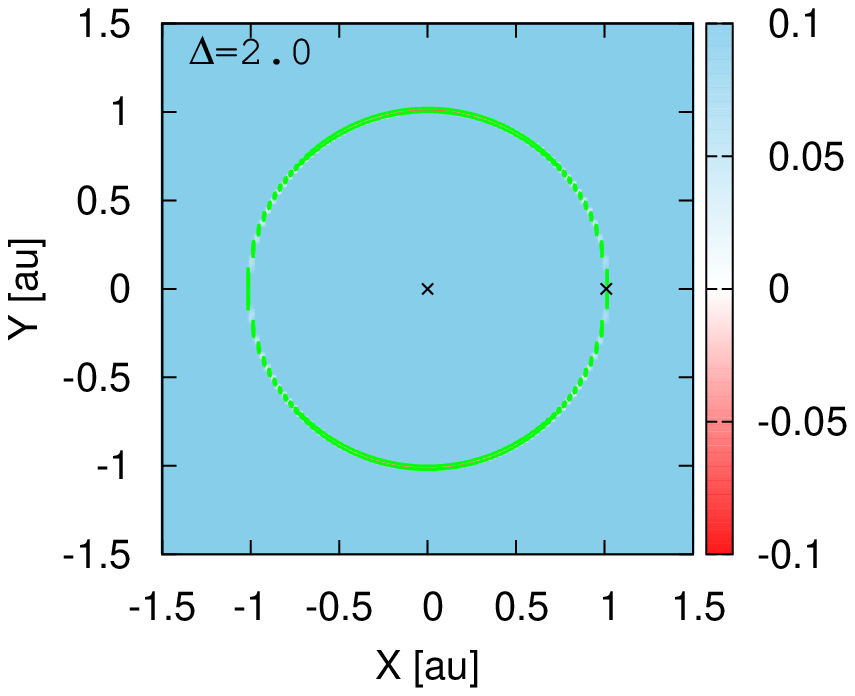}
        \end{minipage}
        \begin{minipage}{8.5cm}
            \includegraphics[width=8.5cm]{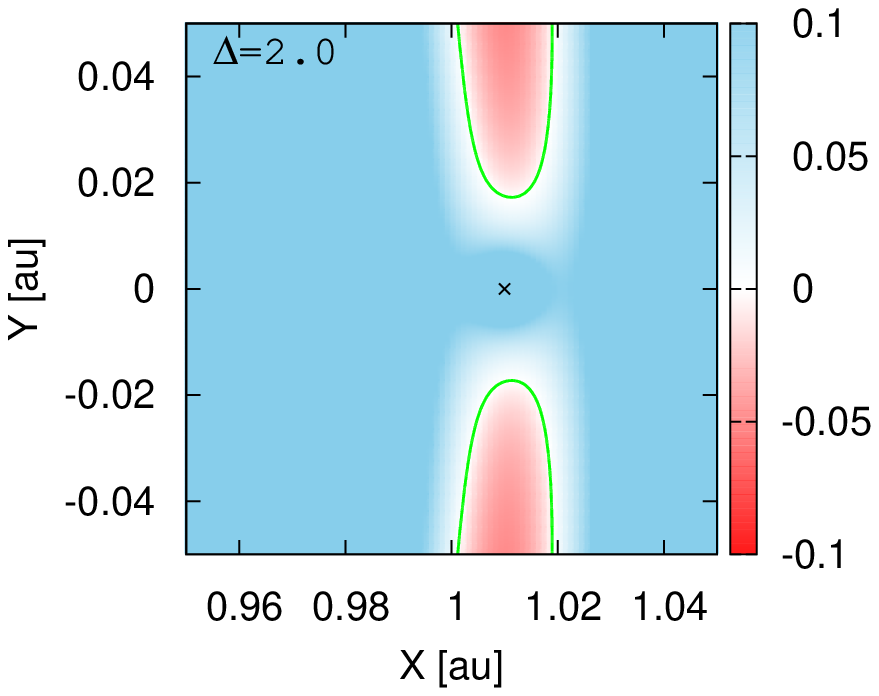}
        \end{minipage}      
        \\
        \begin{minipage}{8.5cm}
            \includegraphics[width=8.5cm]{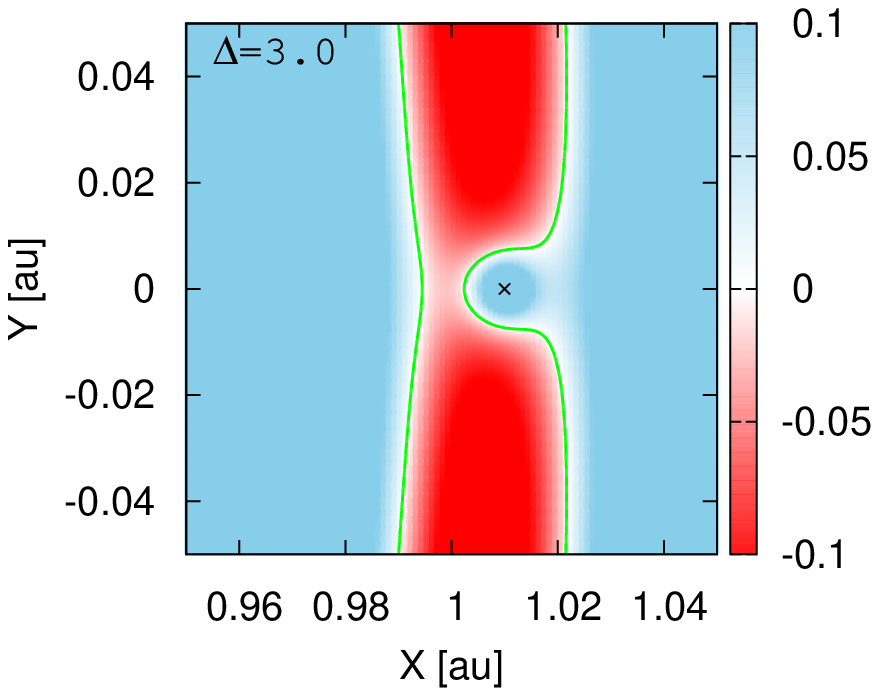}
        \end{minipage} 
        \begin{minipage}{8.5cm}
            \includegraphics[width=8.5cm]{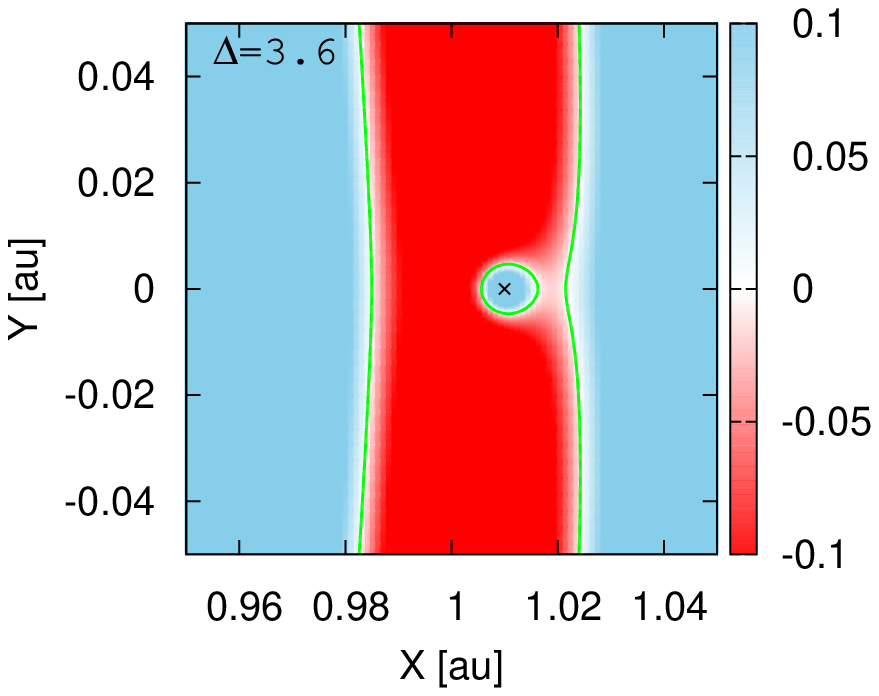}
        \end{minipage} 
        \\
        \caption{
                The Newtonian allowed/forbidden regions for the third body of the circular model in the SMBH group.
                Upper two panels show the results for $\Delta =2.0$. 
                The upper right panel is the zoom-in to the inner orbiting object.
                The counterparts for $\Delta =3.0$ and $3.6$ are displayed in the bottom left and right panels, respectively.
                The value of function $f_\mathrm{N}$ is represented by colors.
                The bluish and reddish regions correspond to the allowed and forbidden regions, respectively.
                The green lines are contours for $f_\mathrm{N}=0$, which are the boundary dividing the two regions.
                Cross points are the positions of the central SMBH and the inner orbiting object. 
                The distance between them is fixed to $1.01a_\mathrm{in}$.
                }
        \label{fig:map_N}
    \end{figure*}
In order to analyze these results further, we map the allowed regions of motion for the third body, using Eq.~\eqref{eq:Sundman3} for the Newtonian and Eq.~\eqref{eq:Sundman_1PN_virial} for the 1PN cases.
In so doing, we need to fix the positions of the central and inner orbiting objects, i.e., $\Delta$ and $r_{12}$, in addition to the values of $\mathcal{H}$ and $J$.
Since $r_{12}$ fluctuates in time as should be obvious from the bottom panel in Fig.~\ref{fig:orbit_ev}, we try a range of values of $r_{12}$.
In Fig.~\ref{fig:map_N}, we show the maps of the allowed regions so drawn for the circular models with $\Delta=2.0$, $3.0$ and $3.6$;
the top left and right panels show the whole map for $\Delta=2.0$ and the zoom-in to the vicinity of the inner orbiting object;
the bottom left and right panels are the zoom-in figures for $\Delta=3.0$ and $\Delta=3.6$, respectively.
The cross points in these figures indicate the positions of the central SMBH and the inner orbiting object.
We fix the value of $r_{12}$ to $1.01a_\mathrm{in}$. 
The color shows the value of 
    \begin{equation}
        f_\mathrm{N} =  2\left( \sum_j m_j r_j^2 \right) \left(  \mathcal{H}_\mathrm{N} - U \right) - J^2.
            \label{eq:f_N}
    \end{equation}
If $f_\mathrm{N}$ is positive (bluish region), inequality~\eqref{eq:Sundman3} is satisfied, that is, the position of concern lies in the allowed region for the third body.
On the other hand, if $f_\mathrm{N}$ is negative (reddish region), the position is in the forbidden region and the third body cannot enter the region.
The green line is a contour for $f_\mathrm{N}=0$, which corresponds to the boundary between the allowed and forbidden regions.

For $\Delta=2.0$, the forbidden region covers the inner orbit except around the inner-orbiting object.
As $\Delta$ increases, the forbidden region is expanded.
As a matter of fact, at $\Delta=3.0$, the forbidden region is extended to the Lagrangian point $L_1$ between the central and inner-orbiting objects;
at $\Delta=3.6$, the forbidden region reaches another Lagrangian point $L_2$ and the inner-orbiting object is now completely surrounded by the forbidden region.
This means that the third body is not allowed to approach the inner-orbiting object as closely as the Hill radius, that is, the system is Hill stable.
This behavior of the Newtonian allowed-region is consistent with what was found by \citet{Marchal82}.

    \begin{figure*}
        \centering
        \begin{minipage}{8.5cm}
            \includegraphics[width=8.5cm]{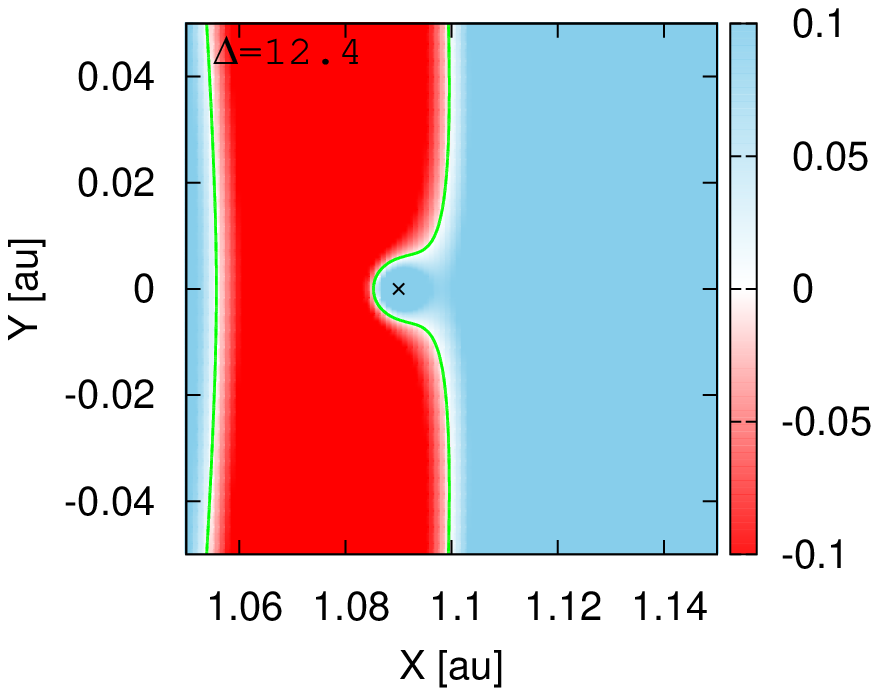}
        \end{minipage} 
        \begin{minipage}{8.5cm}
            \includegraphics[width=8.5cm]{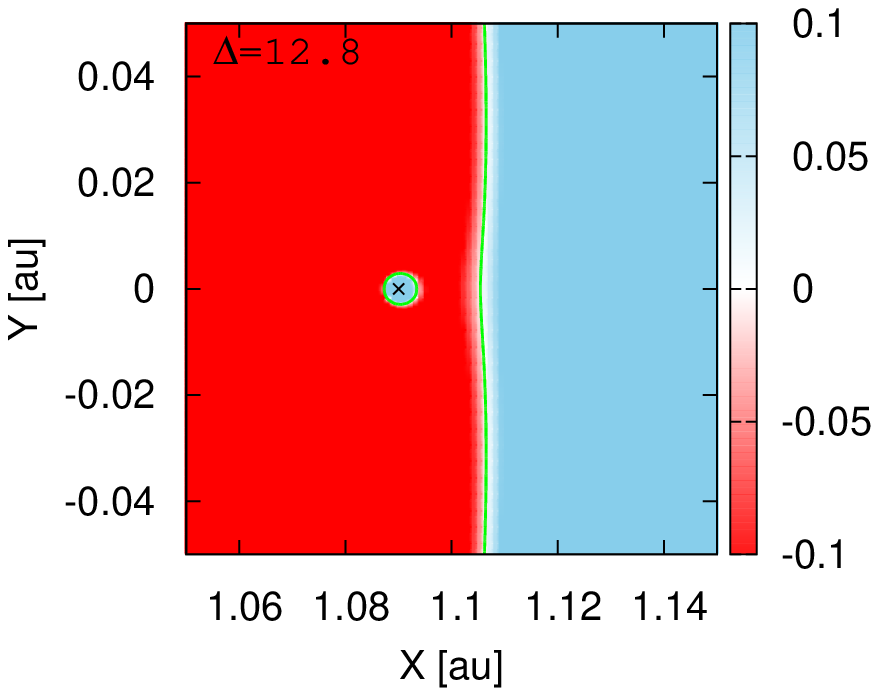}
        \end{minipage} 
        \\
        \caption{
                The 1PN allowed/forbidden regions for the third body of the circular model in the SMBH group.
                The left and right panels show the results for $\Delta = 12.4$ and $12.8$, respectively.
                The value of function $f_\mathrm{1PN}$ is represented by colors.
                The bluish and reddish regions correspond to the allowed and forbidden regions, respectively.
                The green lines are contours for $f_\mathrm{1PN}=0$, which are the boundary dividing the two regions. The cross point in each panel indicates the position of the inner orbiting object. 
                The distance between the central SMBH to the inner orbiting object is fixed to $1.09a_\mathrm{in}$.
                }
        \label{fig:map_PN}
    \end{figure*}
The allowed-region in the 1PN approximation shows a similar behavior.
The zoom-in maps for $\Delta=12.4$ and $12.8$ are exhibited in Fig.~\ref{fig:map_PN}.
The color in this figure shows the value of 
    \begin{eqnarray}
        f_\mathrm{1PN} &=& 2\left( \sum_j m_j r_j^2 \right) 
                            \Bigg\{  
                                    \mathcal{H}_\mathrm{PN} - U 
        \nonumber \\
                                  && + \frac{9}{8} \frac{G^2m_1^2}{c^2}  
                                    \left( 
                                            \frac{m_2}{r_2^2} +  \frac{m_3}{r_3^2} 
                                    \right) 
                            \Bigg\} 
                            - J^2 .
            \label{eq:f_PN}
    \end{eqnarray}
As in the Newtonian maps in Fig.~\ref{fig:map_N}, the bluish and reddish regions correspond to the allowed and forbidden regions for the third body, respectively, and the green line is the boundary between them.

The forbidden region is extended to the inner Lagrangian point $L_1$ at $\Delta=12.4$ (see left panel of Fig.~\ref{fig:map_PN}) whereas at $\Delta=12.8$ it is further expanded to the outer Lagrangian point $L_2$ and covers the inner-orbiting object completely.
These results suggest that the arrival of the forbidden region at $L_2$ may be regarded as the sufficient condition of the Hill stability both in the Newtonian and 1PN calculations.
    \begin{figure*}
        \centering
        \begin{minipage}{8.5cm}
            \includegraphics[width=8.5cm]{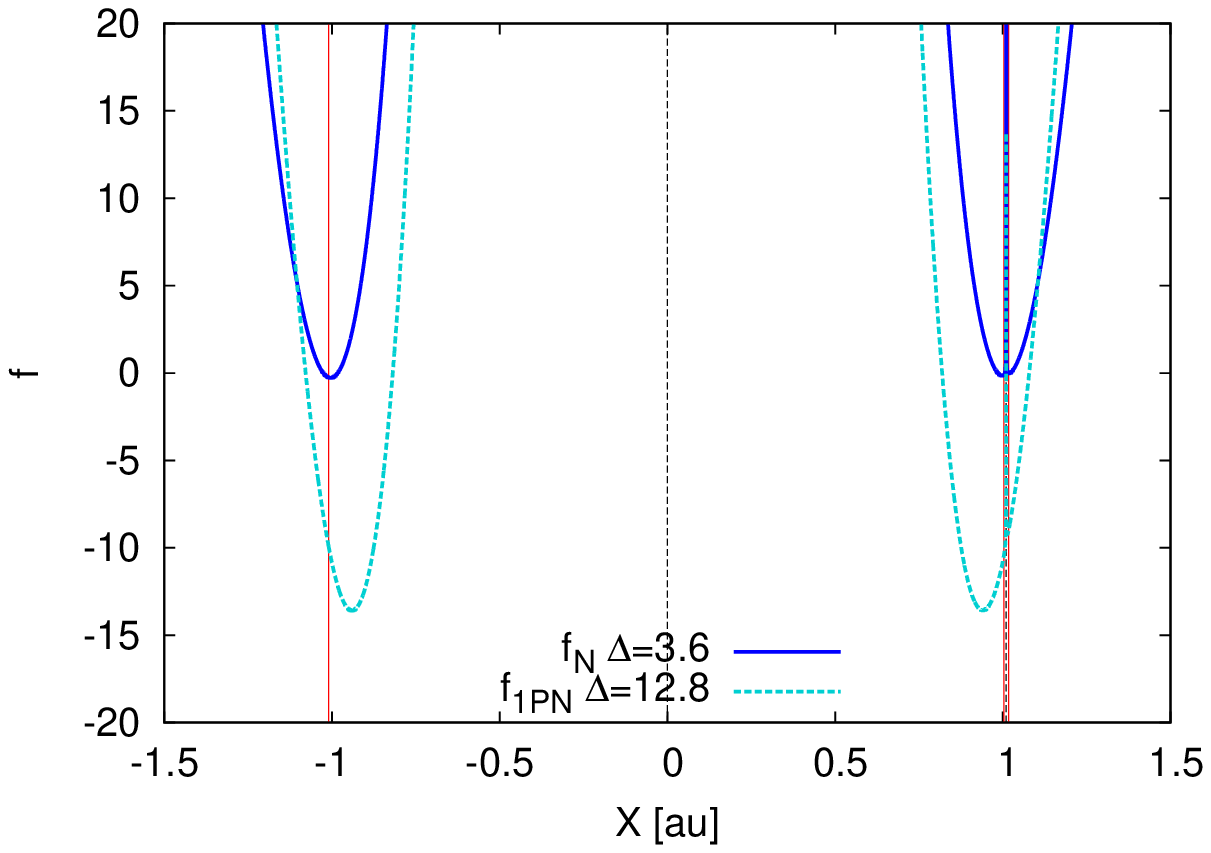}
        \end{minipage}
        \begin{minipage}{8.5cm}
            \includegraphics[width=8.5cm]{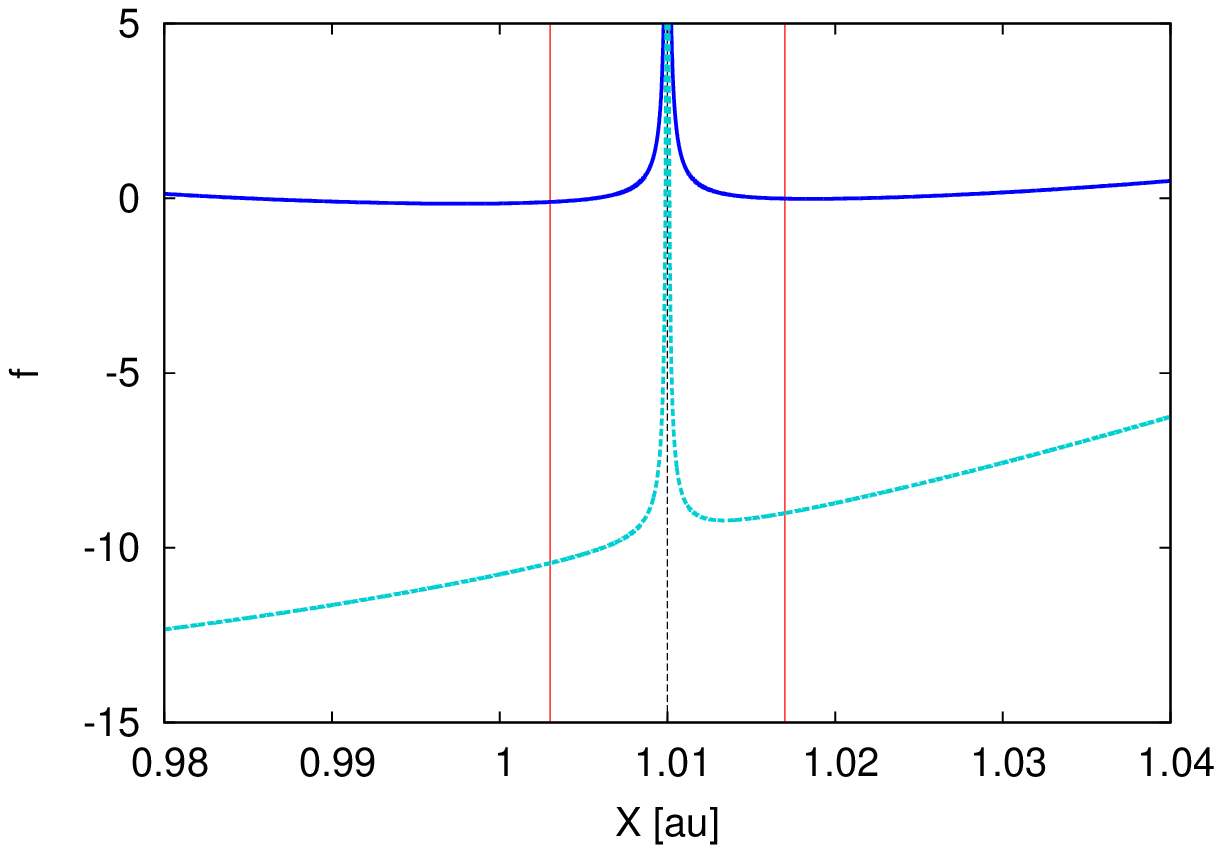}
        \end{minipage}
        \caption{
                The values of functions $f_\mathrm{N}$ (blue) and $f_\mathrm{1PN}$ (cyan) as a function of $X$  on the $Y$ axis.
                The right panel is zoom-in to the inner orbiting object.
                The values of $\Delta$ are $\Delta = 3.6$ and $12.8$ for the Newtonian and 1PN cases, respectively.
                The black lines indicate the positions of the central SMBH and inner orbiting object.
                The distance between them is fixed to $1.01a_\mathrm{in}$ in both calculations.
                The red lines show the Newtonian Lagrangian points.
                }
        \label{fig:fvalue}
    \end{figure*}

The local minimum of $f_\mathrm{1PN}$ is close but not completely identical to the Newtonian counterpart.
Figure~\ref{fig:fvalue} shows the values of $f_\mathrm{N}$ and $f_\mathrm{1PN}$ on the X-axis with $Y=0$ for $\Delta=3.6$ in the Newtonian calculation and for $\Delta = 12.8$ in the 1PN calculation.
The left panel is the whole view and the right one is the zoom-in to the inner orbiting object. 
The blue and cyan lines are the values of $f_\mathrm{N}$ and $f_\mathrm{1PN}$, respectively.
The black dashed lines are the positions of the central and inner orbiting objects.
In this figure, $r_{12}$ is fixed to $1.01a_\mathrm{in}$ in both the Newtonian and 1PN calculations.
The Newtonian Lagrangian points are exhibited as red solid lines;
the X-coordinates of the Newtonian Lagrangian points are given \cite{SSD2000} as
    \begin{eqnarray}
        && X_{L_1} = \Big\{ 1 - \left( \frac{\mu_2}{3} \right)^\frac{1}{3} \Big\} r_{12},\\
        && X_{L_2} = \Big\{ 1 + \left( \frac{\mu_2}{3} \right)^\frac{1}{3} \Big\} r_{12},\\
        && X_{L_3} = -\Big\{ 1 -\frac{7}{12} \mu_2 \Big\} r_{12}.
    \end{eqnarray}
The local minimum points of $f_\mathrm{1PN}$ are slightly dislocated from those of $f_N$, which coincide with the Lagrangian points, are hence referred to as the 1PN Lagrangian points.

The Newtonian and 1PN sufficient conditions are exhibited in Figs.~\ref{fig:SMBH_circular_T} to \ref{fig:SMBH_e_large_T} as the red solid lines.
In drawing these figures, we take the following steps:
fixing $r_{12}$, we first search for the value of $\Delta$ in the range of $R_\mathrm{Hill,2}$ to $50 R_\mathrm{Hill,2}$, at which the forbidden region appears for the first time;
we then vary the value of $r_{12}$ in the range of
$0.5a_\mathrm{in}(e_\mathrm{in}-1)$ to $1.5a_\mathrm{in}(e_\mathrm{in}+1)$, looking for the maximum value of $\Delta$, at which the forbidden region contains the inner-orbiting object.
This value of $\Delta$ is regarded as the sufficient condition of Hill stability.

One finds that the Newtonian red lines agree well with Gladman's sufficient conditions.
As a sufficient condition for the Hill stability, they are indeed consistent with the results of the numerical simulations although they give a bit less tight a criterion for the large-$e$ case.
The discrepancies from Gladman's conditions may be 
due to the fact that we search numerically the maximum $\Delta$ by changing $r_{12}$ and $\Delta$ independently within a finite range.

The 1PN lines, drawn according to our new criterion, on the other hand, are also consistent with the results of the 1PN simulations.
As sufficient conditions for the 1PN Hill stability, it is a little too tight for the circular case as seen in Fig.~\ref{fig:SMBH_circular_T} whereas they are looser for larger eccentricities compared with the Newtonian case.
Some of the approximations in \S\ref{subsec:Hill_PN}
may be responsible for these discrepancies: 
in fact, the virial relation Eq.~\eqref{eq:virial} 
is not strictly satisfied and may have caused the small discrepancy seen in Fig.~\ref{fig:SMBH_circular_T};
in the small- and large-$e$ cases, the approximations used in Eqs.~\eqref{eq:app_r1} and \eqref{eq:app_r2} may be too conservative and may 
have produced the not-so-tight conditions in Figs.~\ref{fig:SMBH_e_small_T} and \ref{fig:SMBH_e_large_T}.

So far we have neglected the higher-order PN terms in our simulations.
Their importance may be roughly estimated as follows.
According to \citet{Barker75}, the timescale $t_\mathrm{LT}$ of the Lense-Thrring precession, which occurs at the 1.5 PN order, is given as
    \begin{eqnarray}
        t_\mathrm{LT} &=& \frac{2 c^3 a_\mathrm{in}^3 (1-e_\mathrm{in}^2)^{3/2} }{ \chi_1 G^2 m_1^2 (4+3m_2/m_1)}
        \nonumber \\
                      &\sim& 1\mathrm{yr} \left( \frac{\chi_1}{0.1} \right)
                                          \left( \frac{a_\mathrm{in}}{1.0\mathrm{au}} \right)^3
                                          \left( \frac{m_1}{10^6 M_\odot} \right)^{-2} ,
    \end{eqnarray}
where $\chi_1 \leq 1$ is the Kerr parameter.
This timescale is rather short and hence 
may have an important effect on the  relativistic Hill stability even if the Kerr parameter of the central SMBH is not so large.
Its detailed analysis will be a future work.

The timescale for GW emissions, which emerge at the 2.5 PN order, is estimated as \cite{Peters64}
    \begin{eqnarray}
        t_\mathrm{GW} &=& \frac{5}{256} \frac{c^5}{G^3}\frac{a_\mathrm{in}}{m_1m_2(m_1+m_2)}
        \nonumber \\
                      &\sim& 10^5 \mathrm{yr} \left( \frac{a_\mathrm{in}}{1.0\mathrm{au}} \right)^4
                                              \left( \frac{m_1}{10^6 M_\odot} \right)^{-2}
                                              \left( \frac{m_2}{1.0 M_\odot} \right)^{-1} .
    \end{eqnarray}
This is essentially the timescale for the merger of the inner orbiting object with the central SMBH, that is, if $T_\mathrm{stab}$ is longer than $t_\mathrm{GW}$, the inner orbit collapses before the system becomes Hill-unstable.
In such a case, the so-called Extreme Mass Ratio Inspiral (EMRI) with an outer perturber might be observed \cite{Amaro-Seoane12, Bonga19}.
How the GW emission affects the sufficient conditions themselves is another interesting topic, which will be addressed in future.

These interesting but unaddressed effects notwithstanding, 
we think that our approximate 1PN sufficient conditions for the relativistic Hill stability is a useful tool, for example, to estimate the stability of multi-body systems, which may be used before conducting costly direct numerical simulations.

\subsection{\label{subsec:res_IMBH} IMBH group}
    \begin{figure}
        \centering
        \includegraphics[width=8.5cm,clip]{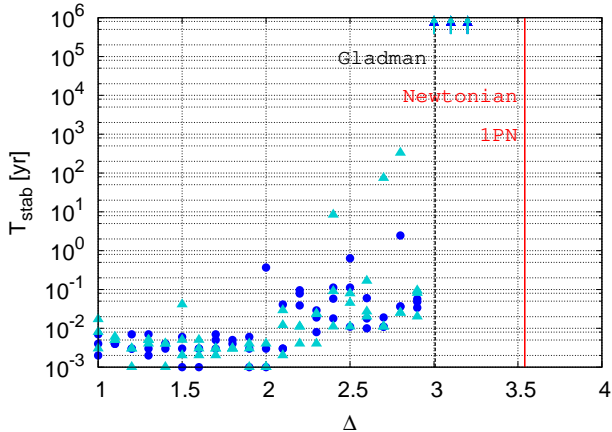}
        \caption{
                The same as Fig.~\ref{fig:SMBH_circular_T} but for the circular model in the IMBH group.
                }
        \label{fig:IMBH_circular_T}
    \end{figure}
    \begin{figure}
        \centering
        \includegraphics[width=8.5cm,clip]{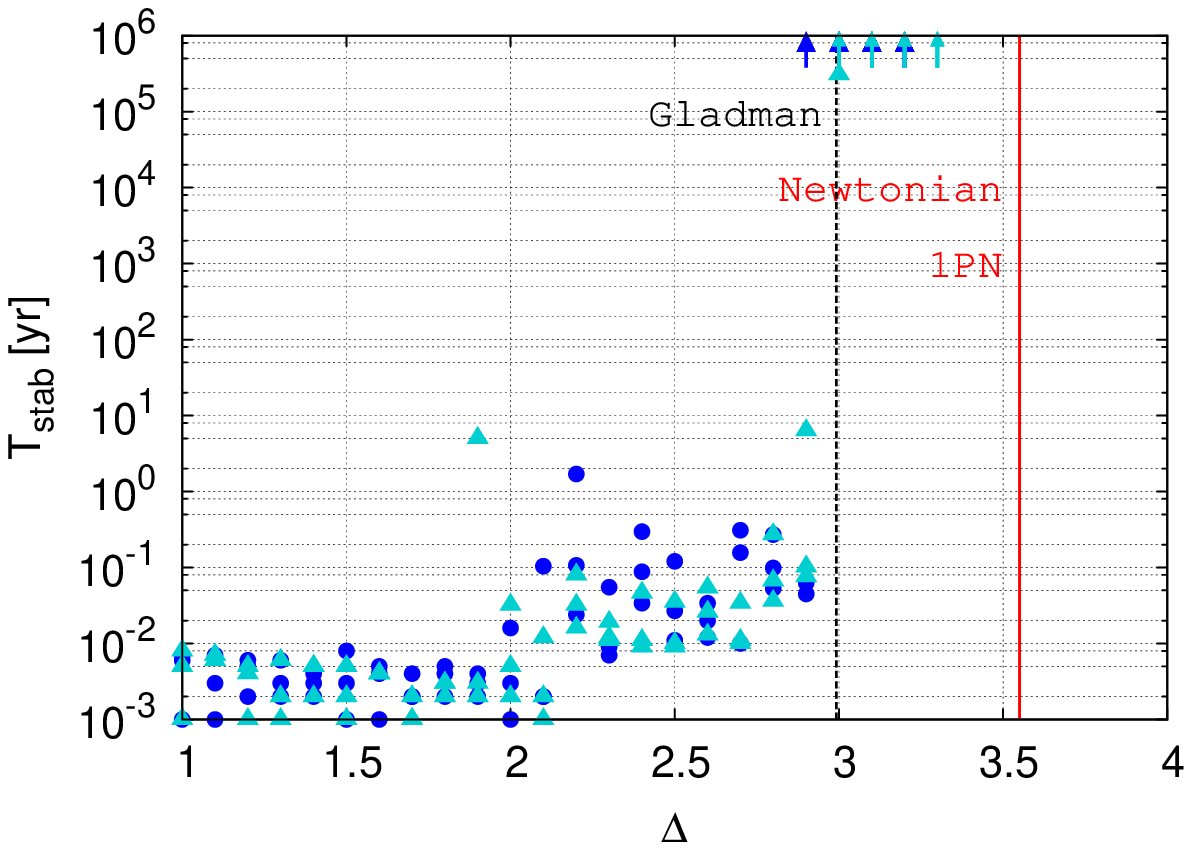}
        \caption{
                The same as Fig.~\ref{fig:SMBH_circular_T} but for the small-$e$ model in the IMBH group.
                }
        \label{fig:IMBH_e_small_T}
    \end{figure}
    \begin{figure}
        \centering
        \includegraphics[width=8.5cm,clip]{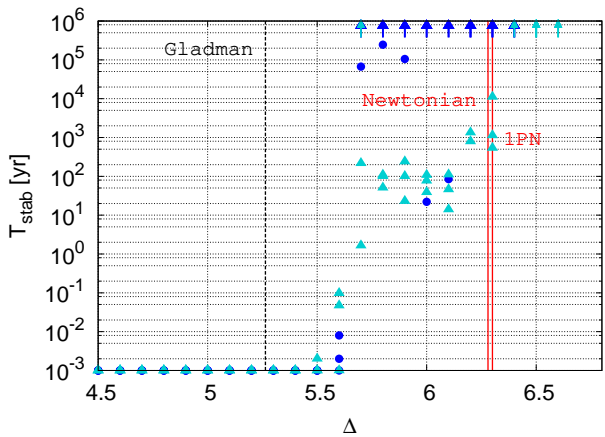}
        \caption{
                The same as Fig.~\ref{fig:SMBH_circular_T} but for the large-$e$ model in the IMBH group.
                }
        \label{fig:IMBH_e_large_T}
    \end{figure}

In the simulations for models in the IMBH group, the relations between $\Delta$ and $T_\mathrm{stab}$ obtained in the 1PN calculations are not so different from the Newtonian ones. 
In fact, they show the same behavior as the Newtonian results of the counterparts in the SMBH group. 
Figures~\ref{fig:IMBH_circular_T} to \ref{fig:IMBH_e_large_T} show the results of the circular, small-$e$ and large-$e$ models, respectively
\footnote{ In our long-term simulations, the typical error in the conservation of total angular momentum is less than $1\%$.
For a small number of models in the IMBH group we found
much larger numerical errors more than a few tens$\%$.
This occurred when large eccentricities are excited in their orbital evolutions.
In these cases, we should have employ much shorter time steps to resolve fast motions near the periastron, which we could not afford, though.
We hence just excluded those apparently failed computations with the relative error more than $5\%$ from the analysis.
}. 

One finds that our Newtonian sufficient conditions are overlapped with the 1PN counterparts.
This is as expected, though, because the last term in the left-hand side of inequality~\eqref{eq:Sundman_1PN_virial} is negligibly small in these cases and inequalities~\eqref{eq:Sundman3} and \eqref{eq:Sundman_1PN_virial} become almost identical.
The timescale of the periastron shift $t_\mathrm{P}$ is
$\sim 2.0\times 10^3$ days in this case and is much longer than $P_\mathrm{in}$ and, more importantly, somewhat longer than $T_\mathrm{stab}$ in the Newtonian case.
This implies again that the 1PN effects do not affect the Hill stability very much.

The 1PN effect on the Hill stability is hence important only for the system with the last term in the left-hand side of Eq.~\eqref{eq:Sundman_1PN_virial} comparable with the total Hamiltonian or the Newtonian potential.
We find that this is not the case for the systems with the $10^3 M_\odot$ IMBH and the inner-orbital semi-major axis $a_\mathrm{in}=0.1$ au. 
The difference between Gladman's sufficient conditions and ours seen in Fig.~\ref{fig:IMBH_circular_T} may be due to our numerical procedure to derive the sufficient conditions from the mapping of the allowed region as we discussed earlier.
It is interesting that Gladman's conditions fail to reproduce the results of our Newtonian simulation in Fig.~\ref{fig:IMBH_circular_T}.
This may be due to some additional approximations used to derive Eq.~\eqref{eq:Gladman_el} such as an expansion in eccentricity, which may not be justified for the large-$e$ model in the IMBH group. 

Finally, we give an estimate of the neglected higher-order PN effects as done in \S\ref{subsec:res_SMBH}.
The timescale of the Lense-Thrring precession, which occurs at the 1.5PN order, is evaluated as
    \begin{equation}
         t_\mathrm{LT} \sim 10^3\mathrm{yr} \left( \frac{\chi_1}{0.1} \right)
                                            \left( \frac{a_\mathrm{in}}{0.1\mathrm{au}} \right)^3
                                            \left( \frac{m_1}{10^3 M_\odot} \right)^{-2}.
    \end{equation}
This is not so long compared with the typical value of $T_\mathrm{stab}$ in Fig.~\ref{fig:IMBH_circular_T} and the 1.5 PN order effect may affect the relativistic Hill stability for the system of current concern.
The secular effect of the GW emission at the 2.5 PN order, on the other hand, is estimated as
    \begin{equation}
        t_\mathrm{GW} \sim 10^7 \mathrm{yr} \left( \frac{a_\mathrm{in}}{0.1\mathrm{au}} \right)^4
                                            \left( \frac{m_1}{10^3 M_\odot} \right)^{-2}
                                            \left( \frac{m_2}{1.0 M_\odot} \right)^{-1} .
    \end{equation}
As discussed in \S\ref{subsec:res_SMBH}, if $T_\mathrm{stab}$ is indeed longer than this value, the inner object will merge with the central object before the system becomes unstable in the sense of Hill stability. 

\section{\label{sec:Conclusion}Conclusion}

We studied the relativistic Hill stability problem for three-body systems containing an SMBH or an IMBH as the central object.
We extended the formalism to obtain the sufficient condition for the Hill stability 
in Newtonian mechanics to relativistic mechanics in the 1PN approximation.
On the theoretical side, we derived approximate sufficient conditions for the relativistic Hill stability 
by substituting the 1PN Hamiltonian and total angular momentum into Sundman's inequality and then employing the virial relation.
We found just as in the Newtonian case that a forbidden region lies between the two orbiting objects in some cases, the fact we adopted to judge Hill stability of the system.

In the numerical analysis, we directly integrated the 1PN equations of motion called the Einstein-Infeld-Hofmann equations with the 6th-order implicit Runge-Kutta method.
Following the previous studies done in Newtonian mechanics, our simulations were conducted for numerous three-body systems with different initial separations $\Delta$ between the orbits to investigate the relation between $\Delta$ and the onset time $T_\mathrm{stab}$ of the orbital instability.

The systems we considered in these simulations were divided into two groups: one containing an SMBH as the central object and the other with an IMBH.
Each group consisted of three models: circular, small-$e$ and large-$e$ models.
The relation between $\Delta$ and $T_\mathrm{stab}$ was investigated for each model in each group, and the result was compared with the sufficient condition derived analytically in this paper.

In the SMBH group, the general relativistic effects are non-negligible.
In fact, the 1PN orbital evolutions were more unstable than the Newtonian counterparts in all models.
The numerical results were consistent with our new criterion as a sufficient condition for the relativistic Hill stability, particularly for the circular and small-$e$ models.
The criterion is not so stringent in the large-$e$ models although it is valid as a sufficient condition.
This is probably because the approximation we adopted for the position $r$ to evaluate the inequality is somewhat too conservative.

In the IMBH group, 
the results obtained in the 1PN calculations are not so different from the Newtonian ones.
This results implies the 1PN effect is not important for the Hill stability of the three-body systems in the IMBH group.
Incidentally, we found that Gladman's conditions are inconsistent with the numerical results for the large-e models whereas our criterion is still valid in these cases.

We estimated but did not include some higher-order PN effects: the Lense-Thirring precession and GW emissions in this paper for simplicity.
The timescales, on which these effects become appreciable, will be short compared with $T_\mathrm{stab}$ near the threshold for stability both in the SMBH and IMBH groups.
This indicates that these processes cannot be ignored to obtain a more tight condition, which will be an interesting topic worth further investigation.
These remaining issues notwithstanding, we think our new conditions will be useful as a measure for the orbital stability of relativistic multi-body systems, which one can employ before conducting costly numerical simulations for such systems.  

\begin{acknowledgments}
We would like to thank Hirotada Okawa and Kensuke Yoshida for the useful discussions.
S. Y. is supported by Institute for Advanced Theoretical and Experimental Physics, and Waseda University and the Waseda University Grant for Special Research Projects (project number: 2020-C273).
This work is supported by JSPS KAKENHI Grant Number JP20J12436.
\end{acknowledgments}

\bibliography{3body_Hill}

\providecommand{\noopsort}[1]{}\providecommand{\singleletter}[1]{#1}%
\begin{thebibliography}{57}%
\makeatletter
\providecommand \@ifxundefined [1]{%
 \@ifx{#1\undefined}
}%
\providecommand \@ifnum [1]{%
 \ifnum #1\expandafter \@firstoftwo
 \else \expandafter \@secondoftwo
 \fi
}%
\providecommand \@ifx [1]{%
 \ifx #1\expandafter \@firstoftwo
 \else \expandafter \@secondoftwo
 \fi
}%
\providecommand \natexlab [1]{#1}%
\providecommand \enquote  [1]{``#1''}%
\providecommand \bibnamefont  [1]{#1}%
\providecommand \bibfnamefont [1]{#1}%
\providecommand \citenamefont [1]{#1}%
\providecommand \href@noop [0]{\@secondoftwo}%
\providecommand \href [0]{\begingroup \@sanitize@url \@href}%
\providecommand \@href[1]{\@@startlink{#1}\@@href}%
\providecommand \@@href[1]{\endgroup#1\@@endlink}%
\providecommand \@sanitize@url [0]{\catcode `\\12\catcode `\$12\catcode
  `\&12\catcode `\#12\catcode `\^12\catcode `\_12\catcode `\%12\relax}%
\providecommand \@@startlink[1]{}%
\providecommand \@@endlink[0]{}%
\providecommand \url  [0]{\begingroup\@sanitize@url \@url }%
\providecommand \@url [1]{\endgroup\@href {#1}{\urlprefix }}%
\providecommand \urlprefix  [0]{URL }%
\providecommand \Eprint [0]{\href }%
\providecommand \doibase [0]{https://doi.org/}%
\providecommand \selectlanguage [0]{\@gobble}%
\providecommand \bibinfo  [0]{\@secondoftwo}%
\providecommand \bibfield  [0]{\@secondoftwo}%
\providecommand \translation [1]{[#1]}%
\providecommand \BibitemOpen [0]{}%
\providecommand \bibitemStop [0]{}%
\providecommand \bibitemNoStop [0]{.\EOS\space}%
\providecommand \EOS [0]{\spacefactor3000\relax}%
\providecommand \BibitemShut  [1]{\csname bibitem#1\endcsname}%
\let\auto@bib@innerbib\@empty
\bibitem [{\citenamefont {Hill}(1878)}]{Hill1878}%
  \BibitemOpen
  \bibfield  {author} {\bibinfo {author} {\bibfnamefont {G.~W.}\ \bibnamefont
  {Hill}},\ }\href@noop {} {\bibfield  {journal} {\bibinfo  {journal} {Amer. J.
  Math.}\ }\textbf {\bibinfo {volume} {1}},\ \bibinfo {pages} {129} (\bibinfo
  {year} {1878})}\BibitemShut {NoStop}%
\bibitem [{\citenamefont {Szebehely}(1967)}]{Szebehely67}%
  \BibitemOpen
  \bibfield  {author} {\bibinfo {author} {\bibfnamefont {V.}~\bibnamefont
  {Szebehely}},\ }\href@noop {} {\emph {\bibinfo {title} {{\it Theory of
  orbits. The restricted problem of three bodies}}}}\ (\bibinfo  {publisher}
  {Academic Press. New York.},\ \bibinfo {year} {1967})\BibitemShut {NoStop}%
\bibitem [{\citenamefont {H\'{e}non}(1970)}]{Henon70}%
  \BibitemOpen
  \bibfield  {author} {\bibinfo {author} {\bibfnamefont {M.}~\bibnamefont
  {H\'{e}non}},\ }\href@noop {} {\bibfield  {journal} {\bibinfo  {journal}
  {Astron. Astrophys.}\ }\textbf {\bibinfo {volume} {9}},\ \bibinfo {pages}
  {24} (\bibinfo {year} {1970})}\BibitemShut {NoStop}%
\bibitem [{\citenamefont {H\'{e}non}\ and\ \citenamefont
  {Petit}(1986)}]{Henon86}%
  \BibitemOpen
  \bibfield  {author} {\bibinfo {author} {\bibfnamefont {M.}~\bibnamefont
  {H\'{e}non}}\ and\ \bibinfo {author} {\bibfnamefont {J.~M.}\ \bibnamefont
  {Petit}},\ }\href@noop {} {\bibfield  {journal} {\bibinfo  {journal}
  {Celestial Mech.}\ }\textbf {\bibinfo {volume} {38}},\ \bibinfo {pages} {67}
  (\bibinfo {year} {1986})}\BibitemShut {NoStop}%
\bibitem [{\citenamefont {Marchal}\ and\ \citenamefont
  {Bozis}(1982)}]{Marchal82}%
  \BibitemOpen
  \bibfield  {author} {\bibinfo {author} {\bibfnamefont {C.}~\bibnamefont
  {Marchal}}\ and\ \bibinfo {author} {\bibfnamefont {G.}~\bibnamefont
  {Bozis}},\ }\href@noop {} {\bibfield  {journal} {\bibinfo  {journal}
  {Celestial Mech.}\ }\textbf {\bibinfo {volume} {26}},\ \bibinfo {pages} {311}
  (\bibinfo {year} {1982})}\BibitemShut {NoStop}%
\bibitem [{\citenamefont {Milani}\ and\ \citenamefont
  {Nobili}(1983)}]{Milani83a}%
  \BibitemOpen
  \bibfield  {author} {\bibinfo {author} {\bibfnamefont {A.}~\bibnamefont
  {Milani}}\ and\ \bibinfo {author} {\bibfnamefont {A.~M.}\ \bibnamefont
  {Nobili}},\ }\href@noop {} {\bibfield  {journal} {\bibinfo  {journal}
  {Celestial Mech.}\ }\textbf {\bibinfo {volume} {31}},\ \bibinfo {pages} {213}
  (\bibinfo {year} {1983})}\BibitemShut {NoStop}%
\bibitem [{\citenamefont {Roy}\ \emph {et~al.}(1984)\citenamefont {Roy},
  \citenamefont {Walker}, \citenamefont {Carusi},\ and\ \citenamefont
  {Valsecchi}}]{Roy84}%
  \BibitemOpen
  \bibfield  {author} {\bibinfo {author} {\bibfnamefont {A.~E.}\ \bibnamefont
  {Roy}}, \bibinfo {author} {\bibfnamefont {I.~W.}\ \bibnamefont {Walker}},
  \bibinfo {author} {\bibfnamefont {A.}~\bibnamefont {Carusi}},\ and\ \bibinfo
  {author} {\bibfnamefont {G.~B.}\ \bibnamefont {Valsecchi}},\ }\href@noop {}
  {\bibfield  {journal} {\bibinfo  {journal} {Astron. Astrophys.}\ }\textbf
  {\bibinfo {volume} {141}},\ \bibinfo {pages} {25} (\bibinfo {year}
  {1984})}\BibitemShut {NoStop}%
\bibitem [{\citenamefont {Wolszczan}\ and\ \citenamefont
  {Frail}(1992)}]{Wolszczan92}%
  \BibitemOpen
  \bibfield  {author} {\bibinfo {author} {\bibfnamefont {A.}~\bibnamefont
  {Wolszczan}}\ and\ \bibinfo {author} {\bibfnamefont {D.}~\bibnamefont
  {Frail}},\ }\href@noop {} {\bibfield  {journal} {\bibinfo  {journal}
  {Nature}\ }\textbf {\bibinfo {volume} {355}},\ \bibinfo {pages} {145}
  (\bibinfo {year} {1992})}\BibitemShut {NoStop}%
\bibitem [{\citenamefont {Gladman}(1993)}]{Gladman93}%
  \BibitemOpen
  \bibfield  {author} {\bibinfo {author} {\bibfnamefont {B.}~\bibnamefont
  {Gladman}},\ }\href@noop {} {\bibfield  {journal} {\bibinfo  {journal}
  {Icarus}\ }\textbf {\bibinfo {volume} {106}},\ \bibinfo {pages} {247}
  (\bibinfo {year} {1993})}\BibitemShut {NoStop}%
\bibitem [{\citenamefont {Chambers}\ \emph {et~al.}(1996)\citenamefont
  {Chambers}, \citenamefont {Wetherill},\ and\ \citenamefont
  {Boss}}]{Chambers96}%
  \BibitemOpen
  \bibfield  {author} {\bibinfo {author} {\bibfnamefont {J.~E.}\ \bibnamefont
  {Chambers}}, \bibinfo {author} {\bibfnamefont {G.~W.}\ \bibnamefont
  {Wetherill}},\ and\ \bibinfo {author} {\bibfnamefont {A.}~\bibnamefont
  {Boss}},\ }\href@noop {} {\bibfield  {journal} {\bibinfo  {journal} {Icarus}\
  }\textbf {\bibinfo {volume} {119}},\ \bibinfo {pages} {261} (\bibinfo {year}
  {1996})}\BibitemShut {NoStop}%
\bibitem [{\citenamefont {Ito}\ and\ \citenamefont {Tanikawa}(1999)}]{Ito99}%
  \BibitemOpen
  \bibfield  {author} {\bibinfo {author} {\bibfnamefont {T.}~\bibnamefont
  {Ito}}\ and\ \bibinfo {author} {\bibfnamefont {K.}~\bibnamefont {Tanikawa}},\
  }\href@noop {} {\bibfield  {journal} {\bibinfo  {journal} {Icarus}\ }\textbf
  {\bibinfo {volume} {139}},\ \bibinfo {pages} {336} (\bibinfo {year}
  {1999})}\BibitemShut {NoStop}%
\bibitem [{\citenamefont {Chatterjee}\ \emph {et~al.}(2008)\citenamefont
  {Chatterjee}, \citenamefont {Ford}, \citenamefont {Matshumura},\ and\
  \citenamefont {Rasio}}]{Chatterjee08}%
  \BibitemOpen
  \bibfield  {author} {\bibinfo {author} {\bibfnamefont {S.}~\bibnamefont
  {Chatterjee}}, \bibinfo {author} {\bibfnamefont {E.~B.}\ \bibnamefont
  {Ford}}, \bibinfo {author} {\bibfnamefont {S.}~\bibnamefont {Matshumura}},\
  and\ \bibinfo {author} {\bibfnamefont {F.~A.}\ \bibnamefont {Rasio}},\
  }\href@noop {} {\bibfield  {journal} {\bibinfo  {journal} {Astrophys. J.}\
  }\textbf {\bibinfo {volume} {686}},\ \bibinfo {pages} {580} (\bibinfo {year}
  {2008})}\BibitemShut {NoStop}%
\bibitem [{\citenamefont {Smith}\ and\ \citenamefont
  {Lissauer}(2009)}]{Smith09}%
  \BibitemOpen
  \bibfield  {author} {\bibinfo {author} {\bibfnamefont {A.~W.}\ \bibnamefont
  {Smith}}\ and\ \bibinfo {author} {\bibfnamefont {J.~J.}\ \bibnamefont
  {Lissauer}},\ }\href@noop {} {\bibfield  {journal} {\bibinfo  {journal}
  {Icurus}\ }\textbf {\bibinfo {volume} {201}},\ \bibinfo {pages} {381}
  (\bibinfo {year} {2009})}\BibitemShut {NoStop}%
\bibitem [{\citenamefont {Pu}\ and\ \citenamefont {Wu}(2015)}]{Pu15}%
  \BibitemOpen
  \bibfield  {author} {\bibinfo {author} {\bibfnamefont {B.}~\bibnamefont
  {Pu}}\ and\ \bibinfo {author} {\bibfnamefont {Y.}~\bibnamefont {Wu}},\
  }\href@noop {} {\bibfield  {journal} {\bibinfo  {journal} {Astrophys. J.}\
  }\textbf {\bibinfo {volume} {807}},\ \bibinfo {pages} {44} (\bibinfo {year}
  {2015})}\BibitemShut {NoStop}%
\bibitem [{\citenamefont {Marzari}\ and\ \citenamefont
  {Weidenschilling}(2002)}]{Marzari02}%
  \BibitemOpen
  \bibfield  {author} {\bibinfo {author} {\bibfnamefont {F.}~\bibnamefont
  {Marzari}}\ and\ \bibinfo {author} {\bibfnamefont {S.~J.}\ \bibnamefont
  {Weidenschilling}},\ }\href@noop {} {\bibfield  {journal} {\bibinfo
  {journal} {Icurus}\ }\textbf {\bibinfo {volume} {156}},\ \bibinfo {pages}
  {570} (\bibinfo {year} {2002})}\BibitemShut {NoStop}%
\bibitem [{\citenamefont {Marzari}(2014)}]{Marzari14}%
  \BibitemOpen
  \bibfield  {author} {\bibinfo {author} {\bibfnamefont {F.}~\bibnamefont
  {Marzari}},\ }\href@noop {} {\bibfield  {journal} {\bibinfo  {journal} {Mon.
  Not. R. Astron. Soc.}\ }\textbf {\bibinfo {volume} {442}},\ \bibinfo {pages}
  {1110} (\bibinfo {year} {2014})}\BibitemShut {NoStop}%
\bibitem [{\citenamefont {Morrison}\ and\ \citenamefont
  {Kratter}(2016)}]{Morrison16}%
  \BibitemOpen
  \bibfield  {author} {\bibinfo {author} {\bibfnamefont {S.~J.}\ \bibnamefont
  {Morrison}}\ and\ \bibinfo {author} {\bibfnamefont {K.~M.}\ \bibnamefont
  {Kratter}},\ }\href@noop {} {\bibfield  {journal} {\bibinfo  {journal}
  {Astrophys. J.}\ }\textbf {\bibinfo {volume} {823}},\ \bibinfo {pages} {118}
  (\bibinfo {year} {2016})}\BibitemShut {NoStop}%
\bibitem [{\citenamefont {Chambers}\ and\ \citenamefont
  {Wetherill}(1998)}]{Chambers98}%
  \BibitemOpen
  \bibfield  {author} {\bibinfo {author} {\bibfnamefont {J.~E.}\ \bibnamefont
  {Chambers}}\ and\ \bibinfo {author} {\bibfnamefont {G.~W.}\ \bibnamefont
  {Wetherill}},\ }\href@noop {} {\bibfield  {journal} {\bibinfo  {journal}
  {Icarus}\ }\textbf {\bibinfo {volume} {136}},\ \bibinfo {pages} {304}
  (\bibinfo {year} {1998})}\BibitemShut {NoStop}%
\bibitem [{\citenamefont {Iwasaki}\ and\ \citenamefont
  {Ohtsuki}(2006)}]{Iwasaki06}%
  \BibitemOpen
  \bibfield  {author} {\bibinfo {author} {\bibfnamefont {K.}~\bibnamefont
  {Iwasaki}}\ and\ \bibinfo {author} {\bibfnamefont {K.}~\bibnamefont
  {Ohtsuki}},\ }\href@noop {} {\bibfield  {journal} {\bibinfo  {journal}
  {Astron. J.}\ }\textbf {\bibinfo {volume} {131}},\ \bibinfo {pages} {3093}
  (\bibinfo {year} {2006})}\BibitemShut {NoStop}%
\bibitem [{\citenamefont {Zhou}\ \emph {et~al.}(2007)\citenamefont {Zhou},
  \citenamefont {Lin},\ and\ \citenamefont {Sun}}]{Zhou07}%
  \BibitemOpen
  \bibfield  {author} {\bibinfo {author} {\bibfnamefont {J.-L.}\ \bibnamefont
  {Zhou}}, \bibinfo {author} {\bibfnamefont {D.~N.~C.}\ \bibnamefont {Lin}},\
  and\ \bibinfo {author} {\bibfnamefont {Y.-S.}\ \bibnamefont {Sun}},\
  }\href@noop {} {\bibfield  {journal} {\bibinfo  {journal} {Astrophys. J.}\
  }\textbf {\bibinfo {volume} {666}},\ \bibinfo {pages} {423} (\bibinfo {year}
  {2007})}\BibitemShut {NoStop}%
\bibitem [{\citenamefont {Ransom}\ \emph {et~al.}(2014)\citenamefont {Ransom},
  \citenamefont {Stairs}, \citenamefont {Archibald}, \citenamefont {Hessels},
  \citenamefont {Kaplan}, \citenamefont {van Kerkwijk}, \citenamefont {Boyles},
  \citenamefont {Deller}, \citenamefont {Chatterjee}, \citenamefont
  {Schechtman-Rook}, \citenamefont {Berndsen}, \citenamefont {Lynch},
  \citenamefont {Lorimer}, \citenamefont {Karako-Argaman}, \citenamefont
  {Kaspi}, \citenamefont {Kondratiev}, \citenamefont {McLaughlin},
  \citenamefont {van Leeuwen}, \citenamefont {Rosen}, \citenamefont {Roberts},\
  and\ \citenamefont {Stovall}}]{Ransom14}%
  \BibitemOpen
  \bibfield  {author} {\bibinfo {author} {\bibfnamefont {S.~M.}\ \bibnamefont
  {Ransom}}, \bibinfo {author} {\bibfnamefont {I.~H.}\ \bibnamefont {Stairs}},
  \bibinfo {author} {\bibfnamefont {A.~M.}\ \bibnamefont {Archibald}}, \bibinfo
  {author} {\bibfnamefont {J.~W.~T.}\ \bibnamefont {Hessels}}, \bibinfo
  {author} {\bibfnamefont {D.~L.}\ \bibnamefont {Kaplan}}, \bibinfo {author}
  {\bibfnamefont {M.~H.}\ \bibnamefont {van Kerkwijk}}, \bibinfo {author}
  {\bibfnamefont {J.}~\bibnamefont {Boyles}}, \bibinfo {author} {\bibfnamefont
  {A.~T.}\ \bibnamefont {Deller}}, \bibinfo {author} {\bibfnamefont
  {S.}~\bibnamefont {Chatterjee}}, \bibinfo {author} {\bibfnamefont
  {A.}~\bibnamefont {Schechtman-Rook}}, \bibinfo {author} {\bibfnamefont
  {A.}~\bibnamefont {Berndsen}}, \bibinfo {author} {\bibfnamefont {R.~S.}\
  \bibnamefont {Lynch}}, \bibinfo {author} {\bibfnamefont {D.~R.}\ \bibnamefont
  {Lorimer}}, \bibinfo {author} {\bibfnamefont {C.}~\bibnamefont
  {Karako-Argaman}}, \bibinfo {author} {\bibfnamefont {V.~M.}\ \bibnamefont
  {Kaspi}}, \bibinfo {author} {\bibfnamefont {V.~I.}\ \bibnamefont
  {Kondratiev}}, \bibinfo {author} {\bibfnamefont {M.~A.}\ \bibnamefont
  {McLaughlin}}, \bibinfo {author} {\bibfnamefont {J.}~\bibnamefont {van
  Leeuwen}}, \bibinfo {author} {\bibfnamefont {R.}~\bibnamefont {Rosen}},
  \bibinfo {author} {\bibfnamefont {M.~S.~E.}\ \bibnamefont {Roberts}},\ and\
  \bibinfo {author} {\bibfnamefont {K.}~\bibnamefont {Stovall}},\ }\href@noop
  {} {\bibfield  {journal} {\bibinfo  {journal} {Nature}\ }\textbf {\bibinfo
  {volume} {505}},\ \bibinfo {pages} {520} (\bibinfo {year}
  {2014})}\BibitemShut {NoStop}%
\bibitem [{\citenamefont {Randall}\ and\ \citenamefont
  {Xianyu}(2019)}]{Randall_2019}%
  \BibitemOpen
  \bibfield  {author} {\bibinfo {author} {\bibfnamefont {L.}~\bibnamefont
  {Randall}}\ and\ \bibinfo {author} {\bibfnamefont {Z.-Z.}\ \bibnamefont
  {Xianyu}},\ }\href@noop {} {\bibfield  {journal} {\bibinfo  {journal}
  {Astrophys. J.}\ }\textbf {\bibinfo {volume} {878}},\ \bibinfo {pages} {75}
  (\bibinfo {year} {2019})}\BibitemShut {NoStop}%
\bibitem [{\citenamefont {Hoang}\ \emph {et~al.}(2019)\citenamefont {Hoang},
  \citenamefont {Naoz}, \citenamefont {Kocsis}, \citenamefont {Will},\ and\
  \citenamefont {Mclver}}]{Hoang19}%
  \BibitemOpen
  \bibfield  {author} {\bibinfo {author} {\bibfnamefont {B.-M.}\ \bibnamefont
  {Hoang}}, \bibinfo {author} {\bibfnamefont {S.}~\bibnamefont {Naoz}},
  \bibinfo {author} {\bibfnamefont {B.}~\bibnamefont {Kocsis}}, \bibinfo
  {author} {\bibfnamefont {M.~F.}\ \bibnamefont {Will}},\ and\ \bibinfo
  {author} {\bibfnamefont {J.}~\bibnamefont {Mclver}},\ }\href@noop {}
  {\bibfield  {journal} {\bibinfo  {journal} {Astrophys. J. Lett.}\ }\textbf
  {\bibinfo {volume} {875}},\ \bibinfo {pages} {L31} (\bibinfo {year}
  {2019})}\BibitemShut {NoStop}%
\bibitem [{\citenamefont {Gupta}\ \emph {et~al.}(2020)\citenamefont {Gupta},
  \citenamefont {Suzuki}, \citenamefont {Okawa},\ and\ \citenamefont
  {Maeda}}]{Gupta19}%
  \BibitemOpen
  \bibfield  {author} {\bibinfo {author} {\bibfnamefont {P.}~\bibnamefont
  {Gupta}}, \bibinfo {author} {\bibfnamefont {H.}~\bibnamefont {Suzuki}},
  \bibinfo {author} {\bibfnamefont {H.}~\bibnamefont {Okawa}},\ and\ \bibinfo
  {author} {\bibfnamefont {K.}~\bibnamefont {Maeda}},\ }\href@noop {}
  {\bibfield  {journal} {\bibinfo  {journal} {Phys. Rev. D}\ }\textbf {\bibinfo
  {volume} {101}},\ \bibinfo {pages} {104053} (\bibinfo {year}
  {2020})}\BibitemShut {NoStop}%
\bibitem [{\citenamefont {Suzuki}\ \emph {et~al.}(2019)\citenamefont {Suzuki},
  \citenamefont {Gupta}, \citenamefont {Okawa},\ and\ \citenamefont
  {Maeda}}]{Suzuki19}%
  \BibitemOpen
  \bibfield  {author} {\bibinfo {author} {\bibfnamefont {H.}~\bibnamefont
  {Suzuki}}, \bibinfo {author} {\bibfnamefont {P.}~\bibnamefont {Gupta}},
  \bibinfo {author} {\bibfnamefont {H.}~\bibnamefont {Okawa}},\ and\ \bibinfo
  {author} {\bibfnamefont {K.}~\bibnamefont {Maeda}},\ }\href@noop {}
  {\bibfield  {journal} {\bibinfo  {journal} {Mon. Not. R. Astron. Soc.}\
  }\textbf {\bibinfo {volume} {486}},\ \bibinfo {pages} {L52} (\bibinfo {year}
  {2019})}\BibitemShut {NoStop}%
\bibitem [{\citenamefont {Suzuki}\ \emph {et~al.}(2020)\citenamefont {Suzuki},
  \citenamefont {Gupta}, \citenamefont {Okawa},\ and\ \citenamefont
  {Maeda}}]{Suzuki20}%
  \BibitemOpen
  \bibfield  {author} {\bibinfo {author} {\bibfnamefont {H.}~\bibnamefont
  {Suzuki}}, \bibinfo {author} {\bibfnamefont {P.}~\bibnamefont {Gupta}},
  \bibinfo {author} {\bibfnamefont {H.}~\bibnamefont {Okawa}},\ and\ \bibinfo
  {author} {\bibfnamefont {K.}~\bibnamefont {Maeda}},\ }\href@noop {}
  {\bibfield  {journal} {\bibinfo  {journal} {Mon. Not. R. Astron. Soc.}\
  }\textbf {\bibinfo {volume} {staa3081}} (\bibinfo {year} {2020})}\BibitemShut
  {NoStop}%
\bibitem [{\citenamefont {Samsing}\ \emph {et~al.}(2014)\citenamefont
  {Samsing}, \citenamefont {MacLeod},\ and\ \citenamefont
  {Ramirez-Ruiz}}]{Samsing_2014}%
  \BibitemOpen
  \bibfield  {author} {\bibinfo {author} {\bibfnamefont {J.}~\bibnamefont
  {Samsing}}, \bibinfo {author} {\bibfnamefont {M.}~\bibnamefont {MacLeod}},\
  and\ \bibinfo {author} {\bibfnamefont {E.}~\bibnamefont {Ramirez-Ruiz}},\
  }\href@noop {} {\bibfield  {journal} {\bibinfo  {journal} {The Astrophysical
  Journal}\ }\textbf {\bibinfo {volume} {784}},\ \bibinfo {pages} {71}
  (\bibinfo {year} {2014})}\BibitemShut {NoStop}%
\bibitem [{\citenamefont {Leigh}\ \emph {et~al.}(2016)\citenamefont {Leigh},
  \citenamefont {Geller},\ and\ \citenamefont {Toonen}}]{Leigh16}%
  \BibitemOpen
  \bibfield  {author} {\bibinfo {author} {\bibfnamefont {N.~W.~C.}\
  \bibnamefont {Leigh}}, \bibinfo {author} {\bibfnamefont {A.~M.}\ \bibnamefont
  {Geller}},\ and\ \bibinfo {author} {\bibfnamefont {S.}~\bibnamefont
  {Toonen}},\ }\href@noop {} {\bibfield  {journal} {\bibinfo  {journal}
  {Astrophys. J.}\ }\textbf {\bibinfo {volume} {818}},\ \bibinfo {pages} {21}
  (\bibinfo {year} {2016})}\BibitemShut {NoStop}%
\bibitem [{\citenamefont {Leigh}\ \emph {et~al.}(2017)\citenamefont {Leigh},
  \citenamefont {Geller}, \citenamefont {McKernan}, \citenamefont {Ford},
  \citenamefont {Mac~Low}, \citenamefont {Bellovary}, \citenamefont {Haiman},
  \citenamefont {Lyra}, \citenamefont {Samsing}, \citenamefont {O'Dowd},
  \citenamefont {Kocsis},\ and\ \citenamefont {Endlich}}]{Leigh17}%
  \BibitemOpen
  \bibfield  {author} {\bibinfo {author} {\bibfnamefont {N.~W.~C.}\
  \bibnamefont {Leigh}}, \bibinfo {author} {\bibfnamefont {A.~M.}\ \bibnamefont
  {Geller}}, \bibinfo {author} {\bibfnamefont {B.}~\bibnamefont {McKernan}},
  \bibinfo {author} {\bibfnamefont {K.~E.~S.}\ \bibnamefont {Ford}}, \bibinfo
  {author} {\bibfnamefont {M.-M.}\ \bibnamefont {Mac~Low}}, \bibinfo {author}
  {\bibfnamefont {J.}~\bibnamefont {Bellovary}}, \bibinfo {author}
  {\bibfnamefont {Z.}~\bibnamefont {Haiman}}, \bibinfo {author} {\bibfnamefont
  {W.}~\bibnamefont {Lyra}}, \bibinfo {author} {\bibfnamefont {J.}~\bibnamefont
  {Samsing}}, \bibinfo {author} {\bibfnamefont {M.}~\bibnamefont {O'Dowd}},
  \bibinfo {author} {\bibfnamefont {B.}~\bibnamefont {Kocsis}},\ and\ \bibinfo
  {author} {\bibfnamefont {S.}~\bibnamefont {Endlich}},\ }\href@noop {}
  {\bibfield  {journal} {\bibinfo  {journal} {Mon. Not. R. Astron. Soc.}\
  }\textbf {\bibinfo {volume} {474}},\ \bibinfo {pages} {5672} (\bibinfo {year}
  {2017})}\BibitemShut {NoStop}%
\bibitem [{\citenamefont {Liu}\ and\ \citenamefont {Lai}(2017)}]{Liu17}%
  \BibitemOpen
  \bibfield  {author} {\bibinfo {author} {\bibfnamefont {B.}~\bibnamefont
  {Liu}}\ and\ \bibinfo {author} {\bibfnamefont {D.}~\bibnamefont {Lai}},\
  }\href@noop {} {\bibfield  {journal} {\bibinfo  {journal} {Astrophys. J.
  Lett.}\ }\textbf {\bibinfo {volume} {846}},\ \bibinfo {pages} {L11} (\bibinfo
  {year} {2017})}\BibitemShut {NoStop}%
\bibitem [{\citenamefont {Zevin}\ \emph {et~al.}(2019)\citenamefont {Zevin},
  \citenamefont {Samsing}, \citenamefont {Rodriguez}, \citenamefont {Haster},\
  and\ \citenamefont {Ramirez-Ruiz}}]{Zevin19}%
  \BibitemOpen
  \bibfield  {author} {\bibinfo {author} {\bibfnamefont {M.}~\bibnamefont
  {Zevin}}, \bibinfo {author} {\bibfnamefont {J.}~\bibnamefont {Samsing}},
  \bibinfo {author} {\bibfnamefont {C.}~\bibnamefont {Rodriguez}}, \bibinfo
  {author} {\bibfnamefont {C.-J.}\ \bibnamefont {Haster}},\ and\ \bibinfo
  {author} {\bibfnamefont {E.}~\bibnamefont {Ramirez-Ruiz}},\ }\href@noop {}
  {\bibfield  {journal} {\bibinfo  {journal} {The Astrophysical Journal}\
  }\textbf {\bibinfo {volume} {871}},\ \bibinfo {pages} {91} (\bibinfo {year}
  {2019})}\BibitemShut {NoStop}%
\bibitem [{\citenamefont {Secunda}\ \emph {et~al.}(2019)\citenamefont
  {Secunda}, \citenamefont {Bellovary}, \citenamefont {Low}, \citenamefont
  {Ford}, \citenamefont {McKernan}, \citenamefont {Leigh}, \citenamefont
  {Lyra},\ and\ \citenamefont {Sándor}}]{Secunda19}%
  \BibitemOpen
  \bibfield  {author} {\bibinfo {author} {\bibfnamefont {A.}~\bibnamefont
  {Secunda}}, \bibinfo {author} {\bibfnamefont {J.}~\bibnamefont {Bellovary}},
  \bibinfo {author} {\bibfnamefont {M.-M.~M.}\ \bibnamefont {Low}}, \bibinfo
  {author} {\bibfnamefont {K.~E.~S.}\ \bibnamefont {Ford}}, \bibinfo {author}
  {\bibfnamefont {B.}~\bibnamefont {McKernan}}, \bibinfo {author}
  {\bibfnamefont {N.~W.~C.}\ \bibnamefont {Leigh}}, \bibinfo {author}
  {\bibfnamefont {W.}~\bibnamefont {Lyra}},\ and\ \bibinfo {author}
  {\bibfnamefont {Z.}~\bibnamefont {Sándor}},\ }\href@noop {} {\bibfield
  {journal} {\bibinfo  {journal} {Astrophys. J.}\ }\textbf {\bibinfo {volume}
  {878}},\ \bibinfo {pages} {85} (\bibinfo {year} {2019})}\BibitemShut
  {NoStop}%
\bibitem [{\citenamefont {Fragione}\ and\ \citenamefont
  {Antonini}(2019)}]{Fragione_Antonini19}%
  \BibitemOpen
  \bibfield  {author} {\bibinfo {author} {\bibfnamefont {G.}~\bibnamefont
  {Fragione}}\ and\ \bibinfo {author} {\bibfnamefont {F.}~\bibnamefont
  {Antonini}},\ }\href@noop {} {\bibfield  {journal} {\bibinfo  {journal} {Mon.
  Not. R. Astron. Soc.}\ }\textbf {\bibinfo {volume} {488}},\ \bibinfo {pages}
  {728} (\bibinfo {year} {2019})}\BibitemShut {NoStop}%
\bibitem [{\citenamefont {Trani}\ \emph
  {et~al.}(2019{\natexlab{a}})\citenamefont {Trani}, \citenamefont {Fujii},\
  and\ \citenamefont {Spera}}]{Trani19a}%
  \BibitemOpen
  \bibfield  {author} {\bibinfo {author} {\bibfnamefont {A.~A.}\ \bibnamefont
  {Trani}}, \bibinfo {author} {\bibfnamefont {M.~S.}\ \bibnamefont {Fujii}},\
  and\ \bibinfo {author} {\bibfnamefont {M.}~\bibnamefont {Spera}},\
  }\href@noop {} {\bibfield  {journal} {\bibinfo  {journal} {Astrophys. J.}\
  }\textbf {\bibinfo {volume} {875}},\ \bibinfo {pages} {42} (\bibinfo {year}
  {2019}{\natexlab{a}})}\BibitemShut {NoStop}%
\bibitem [{\citenamefont {Fragione}\ and\ \citenamefont
  {Bromberg}(2019)}]{Fragione_Bromberg19}%
  \BibitemOpen
  \bibfield  {author} {\bibinfo {author} {\bibfnamefont {G.}~\bibnamefont
  {Fragione}}\ and\ \bibinfo {author} {\bibfnamefont {O.}~\bibnamefont
  {Bromberg}},\ }\href@noop {} {\bibfield  {journal} {\bibinfo  {journal} {Mon.
  Not. R. Astron. Soc.}\ }\textbf {\bibinfo {volume} {488}},\ \bibinfo {pages}
  {4370} (\bibinfo {year} {2019})}\BibitemShut {NoStop}%
\bibitem [{\citenamefont {Trani}\ \emph
  {et~al.}(2019{\natexlab{b}})\citenamefont {Trani}, \citenamefont {Spera},
  \citenamefont {Leigh},\ and\ \citenamefont {Fujii}}]{Trani19b}%
  \BibitemOpen
  \bibfield  {author} {\bibinfo {author} {\bibfnamefont {A.~A.}\ \bibnamefont
  {Trani}}, \bibinfo {author} {\bibfnamefont {M.}~\bibnamefont {Spera}},
  \bibinfo {author} {\bibfnamefont {N.~W.~C.}\ \bibnamefont {Leigh}},\ and\
  \bibinfo {author} {\bibfnamefont {M.~S.}\ \bibnamefont {Fujii}},\ }\href@noop
  {} {\bibfield  {journal} {\bibinfo  {journal} {Astrophys. J.}\ }\textbf
  {\bibinfo {volume} {855}},\ \bibinfo {pages} {135} (\bibinfo {year}
  {2019}{\natexlab{b}})}\BibitemShut {NoStop}%
\bibitem [{\citenamefont {Ge}\ and\ \citenamefont {Alexander}(1991)}]{Ge91}%
  \BibitemOpen
  \bibfield  {author} {\bibinfo {author} {\bibfnamefont {Y.~C.}\ \bibnamefont
  {Ge}}\ and\ \bibinfo {author} {\bibfnamefont {D.}~\bibnamefont {Alexander}},\
  }\href@noop {} {\bibfield  {journal} {\bibinfo  {journal} {Gen. Relativ.
  Gravit.}\ }\textbf {\bibinfo {volume} {23}},\ \bibinfo {pages} {335}
  (\bibinfo {year} {1991})}\BibitemShut {NoStop}%
\bibitem [{\citenamefont {Ge}\ and\ \citenamefont {Leng}(1994)}]{Ge94}%
  \BibitemOpen
  \bibfield  {author} {\bibinfo {author} {\bibfnamefont {Y.~C.}\ \bibnamefont
  {Ge}}\ and\ \bibinfo {author} {\bibfnamefont {X.}~\bibnamefont {Leng}},\
  }\href@noop {} {\bibfield  {journal} {\bibinfo  {journal} {Planet. Space.
  Sci.}\ }\textbf {\bibinfo {volume} {42}},\ \bibinfo {pages} {231} (\bibinfo
  {year} {1994})}\BibitemShut {NoStop}%
\bibitem [{\citenamefont {Einstein}\ \emph {et~al.}(1938)\citenamefont
  {Einstein}, \citenamefont {Infeld},\ and\ \citenamefont {Hoffmann}}]{EIH38}%
  \BibitemOpen
  \bibfield  {author} {\bibinfo {author} {\bibfnamefont {A.}~\bibnamefont
  {Einstein}}, \bibinfo {author} {\bibfnamefont {L.}~\bibnamefont {Infeld}},\
  and\ \bibinfo {author} {\bibfnamefont {B.}~\bibnamefont {Hoffmann}},\
  }\href@noop {} {\bibfield  {journal} {\bibinfo  {journal} {Annals of
  Mathematics. Second series.}\ }\textbf {\bibinfo {volume} {39(1)}},\ \bibinfo
  {pages} {65} (\bibinfo {year} {1938})}\BibitemShut {NoStop}%
\bibitem [{\citenamefont {Richardson}\ and\ \citenamefont
  {Kelly}(1988)}]{Richardson88}%
  \BibitemOpen
  \bibfield  {author} {\bibinfo {author} {\bibfnamefont {D.~L.}\ \bibnamefont
  {Richardson}}\ and\ \bibinfo {author} {\bibfnamefont {T.~J.}\ \bibnamefont
  {Kelly}},\ }\href@noop {} {\bibfield  {journal} {\bibinfo  {journal}
  {Celestial Mech.}\ }\textbf {\bibinfo {volume} {43}},\ \bibinfo {pages} {193}
  (\bibinfo {year} {1988})}\BibitemShut {NoStop}%
\bibitem [{\citenamefont {Chandrasekhar}\ and\ \citenamefont
  {Contopoulos}(1963)}]{Chandrasekhar608}%
  \BibitemOpen
  \bibfield  {author} {\bibinfo {author} {\bibfnamefont {S.}~\bibnamefont
  {Chandrasekhar}}\ and\ \bibinfo {author} {\bibfnamefont {G.}~\bibnamefont
  {Contopoulos}},\ }\bibfield  {title} {\bibinfo {title} {The viral theorem in
  general relativity in the post-newtonian approximation},\ }\href@noop {}
  {\bibfield  {journal} {\bibinfo  {journal} {Proc. Natl. Acad. Sci. U. S. A.}\
  }\textbf {\bibinfo {volume} {49}},\ \bibinfo {pages} {608} (\bibinfo {year}
  {1963})}\BibitemShut {NoStop}%
\bibitem [{\citenamefont {Fang}\ and\ \citenamefont {Huang}(2019)}]{Fang19a}%
  \BibitemOpen
  \bibfield  {author} {\bibinfo {author} {\bibfnamefont {Y.}~\bibnamefont
  {Fang}}\ and\ \bibinfo {author} {\bibfnamefont {Q.-G.}\ \bibnamefont
  {Huang}},\ }\bibfield  {title} {\bibinfo {title} {Secular evolution of
  compact binaries revolving around a spinning massive black hole},\
  }\href@noop {} {\bibfield  {journal} {\bibinfo  {journal} {Phys. Rev. D}\
  }\textbf {\bibinfo {volume} {99}},\ \bibinfo {pages} {103005} (\bibinfo
  {year} {2019})}\BibitemShut {NoStop}%
\bibitem [{\citenamefont {Fang}\ \emph {et~al.}(2019)\citenamefont {Fang},
  \citenamefont {Chen},\ and\ \citenamefont {Huang}}]{Fang19b}%
  \BibitemOpen
  \bibfield  {author} {\bibinfo {author} {\bibfnamefont {Y.}~\bibnamefont
  {Fang}}, \bibinfo {author} {\bibfnamefont {X.}~\bibnamefont {Chen}},\ and\
  \bibinfo {author} {\bibfnamefont {Q.-G.}\ \bibnamefont {Huang}},\ }\href@noop
  {} {\bibfield  {journal} {\bibinfo  {journal} {Astrophys. J.}\ }\textbf
  {\bibinfo {volume} {887}},\ \bibinfo {pages} {210} (\bibinfo {year}
  {2019})}\BibitemShut {NoStop}%
\bibitem [{\citenamefont {Liu}\ \emph {et~al.}(2019)\citenamefont {Liu},
  \citenamefont {Lai},\ and\ \citenamefont {Wang}}]{Liu19}%
  \BibitemOpen
  \bibfield  {author} {\bibinfo {author} {\bibfnamefont {B.}~\bibnamefont
  {Liu}}, \bibinfo {author} {\bibfnamefont {D.}~\bibnamefont {Lai}},\ and\
  \bibinfo {author} {\bibfnamefont {Y.-H.}\ \bibnamefont {Wang}},\ }\href@noop
  {} {\bibfield  {journal} {\bibinfo  {journal} {Astrophys. J.}\ }\textbf
  {\bibinfo {volume} {883}},\ \bibinfo {pages} {L7} (\bibinfo {year}
  {2019})}\BibitemShut {NoStop}%
\bibitem [{\citenamefont {Peters}\ and\ \citenamefont
  {Mathews}(1963)}]{Peters63}%
  \BibitemOpen
  \bibfield  {author} {\bibinfo {author} {\bibfnamefont {P.~C.}\ \bibnamefont
  {Peters}}\ and\ \bibinfo {author} {\bibfnamefont {J.}~\bibnamefont
  {Mathews}},\ }\href@noop {} {\bibfield  {journal} {\bibinfo  {journal} {Phys.
  Rev.}\ }\textbf {\bibinfo {volume} {131}},\ \bibinfo {pages} {435} (\bibinfo
  {year} {1963})}\BibitemShut {NoStop}%
\bibitem [{\citenamefont {Grishin}\ \emph {et~al.}(2017)\citenamefont
  {Grishin}, \citenamefont {Perets}, \citenamefont {Zenati},\ and\
  \citenamefont {Michaely}}]{Grishin17}%
  \BibitemOpen
  \bibfield  {author} {\bibinfo {author} {\bibfnamefont {E.}~\bibnamefont
  {Grishin}}, \bibinfo {author} {\bibfnamefont {H.~B.}\ \bibnamefont {Perets}},
  \bibinfo {author} {\bibfnamefont {Y.}~\bibnamefont {Zenati}},\ and\ \bibinfo
  {author} {\bibfnamefont {E.}~\bibnamefont {Michaely}},\ }\href@noop {}
  {\bibfield  {journal} {\bibinfo  {journal} {Mon. Not. R. Astron. Soc.}\
  }\textbf {\bibinfo {volume} {466}},\ \bibinfo {pages} {276} (\bibinfo {year}
  {2017})}\BibitemShut {NoStop}%
\bibitem [{\citenamefont {Blaes}\ \emph {et~al.}(2002)\citenamefont {Blaes},
  \citenamefont {Lee},\ and\ \citenamefont {Socrates}}]{Blaes02}%
  \BibitemOpen
  \bibfield  {author} {\bibinfo {author} {\bibfnamefont {O.}~\bibnamefont
  {Blaes}}, \bibinfo {author} {\bibfnamefont {M.~H.}\ \bibnamefont {Lee}},\
  and\ \bibinfo {author} {\bibfnamefont {A.}~\bibnamefont {Socrates}},\
  }\href@noop {} {\bibfield  {journal} {\bibinfo  {journal} {Astrophys. J.}\
  }\textbf {\bibinfo {volume} {578}},\ \bibinfo {pages} {775} (\bibinfo {year}
  {2002})}\BibitemShut {NoStop}%
\bibitem [{\citenamefont {Anderson}\ \emph {et~al.}(2017)\citenamefont
  {Anderson}, \citenamefont {Lai},\ and\ \citenamefont {Storch}}]{Anderson17}%
  \BibitemOpen
  \bibfield  {author} {\bibinfo {author} {\bibfnamefont {K.~R.}\ \bibnamefont
  {Anderson}}, \bibinfo {author} {\bibfnamefont {D.}~\bibnamefont {Lai}},\ and\
  \bibinfo {author} {\bibfnamefont {N.~I.}\ \bibnamefont {Storch}},\
  }\href@noop {} {\bibfield  {journal} {\bibinfo  {journal} {Mon. Not. R.
  Astron. Soc.}\ }\textbf {\bibinfo {volume} {467}},\ \bibinfo {pages} {3066}
  (\bibinfo {year} {2017})}\BibitemShut {NoStop}%
\bibitem [{\citenamefont {Murray}\ and\ \citenamefont
  {Dermott}(2000)}]{SSD2000}%
  \BibitemOpen
  \bibfield  {author} {\bibinfo {author} {\bibfnamefont {C.~D.}\ \bibnamefont
  {Murray}}\ and\ \bibinfo {author} {\bibfnamefont {S.~F.}\ \bibnamefont
  {Dermott}},\ }\href@noop {} {\emph {\bibinfo {title} {{\it Solar System
  Dynamics}}}}\ (\bibinfo  {publisher} {Cambridge Univ. Press},\ \bibinfo
  {year} {2000})\BibitemShut {NoStop}%
\bibitem [{\citenamefont {Butcher}(1964)}]{Butcher64}%
  \BibitemOpen
  \bibfield  {author} {\bibinfo {author} {\bibfnamefont {J.~C.}\ \bibnamefont
  {Butcher}},\ }\href@noop {} {\bibfield  {journal} {\bibinfo  {journal} {math.
  Comp.}\ }\textbf {\bibinfo {volume} {18}},\ \bibinfo {pages} {50} (\bibinfo
  {year} {1964})}\BibitemShut {NoStop}%
\bibitem [{\citenamefont {Will}(2014{\natexlab{a}})}]{Will14a}%
  \BibitemOpen
  \bibfield  {author} {\bibinfo {author} {\bibfnamefont {C.~M.}\ \bibnamefont
  {Will}},\ }\href@noop {} {\bibfield  {journal} {\bibinfo  {journal} {Phys.
  Rev. D}\ }\textbf {\bibinfo {volume} {{89}}},\ \bibinfo {pages} {044043}
  (\bibinfo {year} {2014}{\natexlab{a}})}\BibitemShut {NoStop}%
\bibitem [{\citenamefont {Will}(2014{\natexlab{b}})}]{Will14b}%
  \BibitemOpen
  \bibfield  {author} {\bibinfo {author} {\bibfnamefont {C.~M.}\ \bibnamefont
  {Will}},\ }\href@noop {} {\bibfield  {journal} {\bibinfo  {journal} {Class.
  Quantum Gravity}\ }\textbf {\bibinfo {volume} {{31}}},\ \bibinfo {pages}
  {244001} (\bibinfo {year} {2014}{\natexlab{b}})}\BibitemShut {NoStop}%
\bibitem [{\citenamefont {Migaszewski}\ and\ \citenamefont
  {Go\'{z}dziewski}(2011)}]{Migaszewski11}%
  \BibitemOpen
  \bibfield  {author} {\bibinfo {author} {\bibfnamefont {C.}~\bibnamefont
  {Migaszewski}}\ and\ \bibinfo {author} {\bibfnamefont {K.}~\bibnamefont
  {Go\'{z}dziewski}},\ }\href@noop {} {\bibfield  {journal} {\bibinfo
  {journal} {Mon. Not. R. Astron. Soc.}\ }\textbf {\bibinfo {volume} {411}},\
  \bibinfo {pages} {565} (\bibinfo {year} {2011})}\BibitemShut {NoStop}%
\bibitem [{\citenamefont {Barker}\ and\ \citenamefont
  {O'Connell}(1975)}]{Barker75}%
  \BibitemOpen
  \bibfield  {author} {\bibinfo {author} {\bibfnamefont {B.~M.}\ \bibnamefont
  {Barker}}\ and\ \bibinfo {author} {\bibfnamefont {R.~F.}\ \bibnamefont
  {O'Connell}},\ }\href@noop {} {\bibfield  {journal} {\bibinfo  {journal}
  {Phys. Rev. D}\ }\textbf {\bibinfo {volume} {12}},\ \bibinfo {pages} {329}
  (\bibinfo {year} {1975})}\BibitemShut {NoStop}%
\bibitem [{\citenamefont {Peters}(1964)}]{Peters64}%
  \BibitemOpen
  \bibfield  {author} {\bibinfo {author} {\bibfnamefont {P.~C.}\ \bibnamefont
  {Peters}},\ }\href@noop {} {\bibfield  {journal} {\bibinfo  {journal} {Phys.
  Rev.}\ }\textbf {\bibinfo {volume} {136}},\ \bibinfo {pages} {B1224}
  (\bibinfo {year} {1964})}\BibitemShut {NoStop}%
\bibitem [{\citenamefont {Amaro-Seoane}\ \emph {et~al.}(2012)\citenamefont
  {Amaro-Seoane}, \citenamefont {Brem}, \citenamefont {Cuadra},\ and\
  \citenamefont {Armitage}}]{Amaro-Seoane12}%
  \BibitemOpen
  \bibfield  {author} {\bibinfo {author} {\bibfnamefont {P.}~\bibnamefont
  {Amaro-Seoane}}, \bibinfo {author} {\bibfnamefont {P.}~\bibnamefont {Brem}},
  \bibinfo {author} {\bibfnamefont {J.}~\bibnamefont {Cuadra}},\ and\ \bibinfo
  {author} {\bibfnamefont {P.~J.}\ \bibnamefont {Armitage}},\ }\href@noop {}
  {\bibfield  {journal} {\bibinfo  {journal} {Astrophys. J. Lett.}\ }\textbf
  {\bibinfo {volume} {744}},\ \bibinfo {pages} {L20} (\bibinfo {year}
  {2012})}\BibitemShut {NoStop}%
\bibitem [{\citenamefont {Bonga}\ \emph {et~al.}(2019)\citenamefont {Bonga},
  \citenamefont {Yang},\ and\ \citenamefont {Hughes}}]{Bonga19}%
  \BibitemOpen
  \bibfield  {author} {\bibinfo {author} {\bibfnamefont {B.}~\bibnamefont
  {Bonga}}, \bibinfo {author} {\bibfnamefont {H.}~\bibnamefont {Yang}},\ and\
  \bibinfo {author} {\bibfnamefont {S.~A.}\ \bibnamefont {Hughes}},\
  }\href@noop {} {\bibfield  {journal} {\bibinfo  {journal} {Phys. Rev. Lett.}\
  }\textbf {\bibinfo {volume} {123}},\ \bibinfo {pages} {101103} (\bibinfo
  {year} {2019})}\BibitemShut {NoStop}%
\end{thebibliography}%

\appendix
\section{\label{app:Sundman} Proof of Sundman's inequality}
The Sundman's inequality \eqref{eq:Sundman} can be proved by using the well-known Cauchy's inequality, which is given as
    \begin{equation}
        \left| \sum_j (A_j B_j) \right|^2 \leq \left| \sum_j (A_j)^2 \right| \left| \sum_j  (B_j)^2 \right|, 
    \end{equation}
where $A_j$ and $B_j$ are the components of arbitrary vectors $\bm{A}$ and $\bm{B}$.
The components in right-hand side of Eq.\eqref{eq:Sundman} are estimated as
    \begin{eqnarray}
        \left| \sum_j m_j \bm{r}_j \times \bm{v}_j \right| 
            &\leq& \sum_j m_j r_j v_j  \left| \sin \gamma_j \right|
        \nonumber \\
            &=& \sum_j \sqrt{m_j r_j^2} \sqrt{m_j v_j^2 \sin^2 \gamma_j}
        \label{eq:vector product}
    \end{eqnarray}
    \begin{eqnarray}
        \left| \sum_j m_j \bm{r}_j \cdot \bm{v}_j \right|
            &\leq&  \sum_j m_j r_j v_j \left| \cos \gamma_j \right|
        \nonumber \\
            &=& \sum_j \sqrt{m_j r_j^2} \sqrt{m_j v_j^2 \cos^2 \gamma_j},
        \label{eq:inner product}
    \end{eqnarray}
where $\gamma_j$ is the angle between $\bm{r}_j$ and $\bm{v}_j$.
Applying Cauchy's inequality to the square value of Eq.\eqref{eq:vector product} and \eqref{eq:inner product} gives 
    \begin{eqnarray}
            && \left| \sum_j m_j \bm{r}_j \times \bm{v}_j \right|^2 
                   \leq \left| \sum_j \sqrt{m_j r_j^2} \sqrt{m_j v_j^2 \sin^2 \gamma_j} \right|^2 
            \nonumber \\
                && \leq \left( \sum_j m_j r_j^2  \right) \left( \sum_j  m_j v_j^2 \sin^2 \gamma_j \right) ,
    \end{eqnarray}
    \begin{eqnarray}
            && \left| \sum_j m_j \bm{r}_j \cdot \bm{v}_j \right|^2 
                    \leq \left| \sum_j \sqrt{m_j r_j^2} \sqrt{m_j v_j^2 \cos^2 \gamma_j} \right|^2
            \nonumber \\
                &&  \leq \left( \sum_j m_j r_j^2  \right) \left( \sum_j  m_j v_j^2 \cos^2 \gamma_j \right) .
    \end{eqnarray}
Sundman's inequality \eqref{eq:Sundman} is immediately obtained if the summation of both two inequalities are taken.

\section{\label{app:transformation}Transformation of Orbital Elements}
\subsection{\label{app:initial}Initial Condition}
Initial configurations of our models are set up by using six orbital elements:
semi-major axis $a$, eccentricity $e$, inclination $i$, argument of periastron $\omega$, longitude of ascending node $\Omega$, and mean anomaly $M$.
These orbital elements are transformed to the Cartesian coordinates of the constituent bodies. 
Here we describe the transformation of orbital elements assuming a general orbit that has $i \neq 0$ and $e \neq 0$, which means its longitude of ascending node $\Omega$ and argument of periastron $\omega$ can be defined.
In case of orbit with $i = 0$, which means longitude of ascending node $\Omega$ cannot be defined, $\Omega$ in below equations can be neglected.
In case of circular orbit with $e = 0$, which means the argument of periaston $\omega$ cannot be defined, we alternatively have another degree of freedom to fix the x-axis in the Cartesian coordinates. 
More detail explanations about orbital elements are in \cite{SSD2000}, for example.

First, we calculate the eccentric anomaly $u$ by solving the following equation with the Newton-Raphsom method: 
    \begin{equation}
        M = u-e \sin u .
    \end{equation}
We transform $u$ to the true anomaly $\nu$ with the following equation, 
    \begin{equation}
        \nu = \arctan \left\{ \frac{\sin u \sqrt{1 - e ^{2}}}{\cos u - e} \right\} .
    \end{equation}
The true anomaly $\nu$ gives the polar coordinates of a body on the orbit as
    \begin{eqnarray}
        r &=& \frac{a(1-e^{2})}{1-e \cos \nu} ,\\
        \psi &=& \Omega +\arctan\{ \tan (\omega + \nu) \cos i) \} ,\\
        \theta &=& \arccos\{ \sin(\omega + \nu) \sin i \}.  
    \end{eqnarray}
The origin of these coordinates is put at the position of the central star in our models.
The velocity of a body in these coordinates is described as
    \begin{eqnarray}
        \dot{r}&=&g_{r} \dot{\nu} ,\\
        \dot{\theta}&=&g_{\theta} \dot{\nu} , \\
        \dot{\psi}&=&g_{\psi} \dot{\nu} ,
    \end{eqnarray}
 where $g_{r}$, $g_{\theta}$, $g_{\psi}$, and $\dot{\nu}$ are given as 
    \begin{eqnarray}
      g_{r} &=& \frac{a(1- e^{2}) e \sin \nu}{(1+ e \cos \nu) ^{2}} ,\\
      g_{\theta}&=& - \frac{1}{\sin\theta} \cos{(\omega + \nu)} \sin i ,\\
      g_{\psi}&=& \cos^{2}(\psi-\Omega ) \frac{\cos i}{\cos^{2}(\omega + \nu )} ,
    \end{eqnarray}
    \begin{equation}
     \dot{\nu} = \sqrt{G (m_1+m_i) \left(\frac{2}{r}-\frac{1}{a} \right)
                 \frac{1}{f_{r}^{2} +(r f_{\theta})^{2} +(r \sin \theta f_{\psi})^{2}}} . 
    \end{equation}
We then change these polar coordinates to the Cartesian coordinates and shift their origins to the center of the mass of the entire system 
The numerical integration are done on these Cartesian coordinates.

\subsection{\label{app:post}post-Process}
The computational results are transformed back to the orbital elements of the {\it osculating orbit} of each timestep.
Here we explain the way to get all six Kepler elements from the instantaneous position and velocity. 
We remark that although what we especially need in this paper is only the semi-major axis and eccentricity, the other orbital elements, for example the inclination, will be important in more general analysis that will be done in future work.
The semi-major axis $a$ is obtained as,
    \begin{equation}
      a=-\frac{G (m_1+m_i)}{2 E} .          
      \label{semi-major}
    \end{equation}
In this expression, 
$E$ is the specific orbital energy given as
    \begin{equation}
      E=\frac{1}{2}v^{2}-\frac{G(m_1+m_i)}{r} ,    
    \end{equation}
where $v$ and $r$ are the absolute values of relative velocity $\bm{v}=\bm{v}_i-\bm{v}_1$ and relative position vector $\bm{r}=\bm{x}_i-\bm{x}_1$.
The inclination $i$, eccentricity $e$, and longitude of the ascending node $\Omega$ are described as the following equations:
   \begin{eqnarray}
        i &=& \arccos \left( \frac{(\bm{r} \times \bm{v})_{z}}{|\bm{r} \times \bm{v}|} \right)  ,
        \\
        e &=& \sqrt{1-\frac{|\bm{r} \times \bm{v}|^{2}}{a G (m_1+m_i)}} ,
        \\
        \Omega &=& \arccos \left( \frac{(\bm{n} \times ( \bm{r} \times\bm{v} ))_{x}}
                                            {|\bm{n} \times ( \bm{r} \times \bm{v} )|} \right) ,
    \end{eqnarray}
where the subscripts stand for the components of vectors and $\bm{n}$ is the unit vector normal to the x-y plane of the reference frame.
The argument of periastron $ \omega $ is obtained as following way.
At first, the true anomaly $f$ is calculated as 
    \begin{equation}
        f = \arccos \left( \frac{a(1-e^{2})-r}{e r} \right) .
    \end{equation}
Next, the angle of the orbiting object from the ascending node on the orbital plain $\theta$ is also calculated as
    \begin{equation}
        \theta = \arccos \left( \frac{x\cos{\Omega} + y\sin{\Omega}}{r} \right).
    \end{equation} 
The argument of periastron is finally obtained as the difference of these arguments,
\begin{equation}
 \omega = \theta - f .
\end{equation} 

\end{document}